\documentclass[12pt,citesort]{iopart}

 \expandafter\let\csname equation*\endcsname\relax
 \expandafter\let\csname endequation*\endcsname\relax
 \usepackage{amsmath}
\usepackage{amssymb}
\usepackage{iopams}
\usepackage{graphicx,epsfig,epstopdf}
\usepackage[breaklinks=true,colorlinks=true,linkcolor=blue,urlcolor=blue,citecolor=blue]{hyperref}

\begin{document}

\title{Multidimensional spectroscopy with entangled light;\\loop vs ladder delay scanning protocols.}

\author{Konstantin E. Dorfman$^{*}$ and Shaul Mukamel$^{\dagger}$}
\address{Department of Chemistry, University of California, Irvine,
California 92697-2025, USA}

\ead{$^{*}$kdorfman@uci.edu and $^{\dagger}$smukamel@uci.edu}

\begin{abstract}
Multidimensional optical signals are commonly recorded by varying the delays between time ordered pulses. These control the evolution of the density matrix and are described by ladder diagrams. We propose a new non-time-ordered protocol based on following the time evolution of the wavefunction and described by loop diagrams. The time variables in this protocol allow to observe different types of resonances and reveal information about intraband dephasing not readily available  by time ordered techniques. The time variables involved in this protocol become coupled when using entangled light, which provides high selectivity and background free measurement of the various resonances. Entangled light can resolve certain states even when strong background due to fast  dephasing suppresses the resonant features when probed by classical light.
\end{abstract}

\pacs{42.62.Fi, 78.47.J-, 42.65.Lm}
\maketitle

\tableofcontents


\section{Introduction}

In coherent nonlinear optical spectroscopy the applied optical pulses induce a polarization in the matter system which is then measured. There are  two types of bookkeeping representations for computing an observable (such as the polarization) in a quantum  system subjected to time dependent perturbations. Both are exact and should yield the same final results provided no approximations are made. However they offer a very different physical picture and suggest different types of approximations  that lead to different predictions. 

In the first representation we follow the evolving density matrix in real time. This representation is  most suitable  for impulsive  experiments involving  sequences of short, temporally well-separated, pulses ranging from NMR to the X-ray regimes \cite{Muk95}. The time variables used to represent the delays  between successive pulses  \cite{Abr09} $t_1$, $t_2$, $t_3$, ... serve as the primary control parameters. Spectra are displayed vs the Fourier conjugates $\tilde{\Omega}_1$, $\tilde{\Omega}_2$, $\tilde{\Omega}_3$, ... to these variables. Such signals can be represented by ladder diagrams (see Fig. 1b and Fig. 2).  We shall denote this way of displaying the multidimensional signals as the ladder delay scanning protocol (LAP).  The signals with different phase matching directions are distinct when displayed vs ladder delays. The density matrix further allows for  reduced  descriptions where bath degrees of freedom which cause pure dephasing and relaxation processes are eliminated. 

Alternatively we can follow the evolving wave function. Rather than keeping track of both the bra and the ket we can place the entire burden of the time
evolution on the ket. In that case we must use artificial time variables  where the ket first evolves forward and then backward in time, eventually 
returning to the initial time. This is represented by loop diagrams \cite{Rah10} as is commonly done in many body theory \cite{Ram07}. This gives more compact description (fewer terms). It is harder to visualize
impulsive experiments in this language. However it proves most useful for frequency domain techniques involving long pulses where the time evolution is not monitored
directly \cite{Rah10}. In this picture we give up the full control over time ordering between pulses. We will denote the delays along the loop as $\tau_1$, $\tau_2$, $\tau_3$, ... (see Fig. 1a,c). By displaying the spectra vs the Fourier conjugates to the loop times $\Omega_1$, $\Omega_2$, $\Omega_3$ we obtain the loop delay scanning protocol (LOP).

In  this paper we compare the two display protocols for multidimensional spectroscopy in molecular aggregates  with fluorescence detection. Since the two protocols use different time variables the resulting multidimensional signals obtained by Fourier  transforms conjugate to these variables
appear very different and highlight different resonances. This can be exploited  for  highlighting desired features in optical signals. We  further show  some advantages of the loop representation for describing measurements with quantum light, i.e. entangled broadband photons which have intermediate time/frequency character. We should emphasize that
these protocols offer two languages  for describing the same physics. However the translation is  somewhat tricky making them suitable  for  different applications. We show  how such LOP signals can be realized experimentally and compare it to the LAP.  

The utility of each protocol depends on experimental details including e.g. the system dynamics, bath effects and the specific light field configuration.  For instance when the system is in a pure state and the fields are classical, the loop delays $\tau_{j}$, $j=1,2,3$ which represent forward and backward time propagation periods of the wave function  are the natural independent variables and it makes sense to adopt their conjugate frequencies for display, thus using the LOP. If pure dephasing processes due to a bath are added the signal may no longer factorize into a product of terms each depending on a single delay $\tau_{j}$ when calculating the optical response. In this case the ladder variables $t_{j}$ which represent the LAP delays in real time and correspond to propagation of a density matrix become more natural since the signal can be recast as a product of individual terms each depending on a single $t_{j}$ variable. Stochastic or entangled light fields cause additional coupling between the interaction times imposing that the signal may not generally be factorized in either protocol since the field correlation functions depend on products of factors that depend on pairs of times. In that case neither protocol allows the observed signals to be factorized in a simple way discussed above. The two protocols  highlight different resonances and processes. In the following we demonstrate  what type of information can be extracted from each protocol for Frenkel excitons in a model molecular aggregate.

We further compare signals obtained with classical vs quantum light (entangled photons). The LAP and LOP denote the protocols for displaying multidimensional signals. Calculations performed with either the wavefunction or the density matrix can be displayed using either protocol. In earlier studies ladder diagrams were denoted as double-sided Feynman diagrams, and loop diagrams were denoted as close-time-path-loops (CTPL) \cite{Rah10}.

We investigate the multidimensional signals in a molecular aggregate obtained by incoherent two-photon absorption (TPA) detection. Incoherent detection is often more sensitive than heterodyne as the latter is limited by the pulse duration so there are fewer constraints on the laser system. In addition the low intensity requirements for biological samples limit the range of heterodyne detection setups. This have been demonstrated \cite{Tek07,War07,War09} even in single molecule spectroscopy \cite{Bri10}. Historically Ramsey fringes  constitute the first example of incoherent detection \cite{Ram50,Coh11,Sch131}. Information similar to coherent spectroscopy can be extracted from the parametric dependence on various pulse sequences applied prior to the incoherent detection \cite{Ric11,Rah101}. Possible incoherent detection modes include fluorescence \cite{Moe03, Elf07, Bar08}, photoaccoustic \cite{Pat81,Wan11,Wan13}, AFM \cite{Mam05,Pog09,Raj10,Raj11} or photocurrent detection \cite{Che13,Nar13}. 

Quantum spectroscopy which utilizes the quantum nature of light to reveal matter properties is an emerging field. Entangled photons is one notable example and offer several advantages. First, the signals scale to lower order in the incoming intensity \cite{Lee06}. The pump-probe signal e.g. scales linearly rather than quadratically. This allows to to perform nonlinear spectroscopy with much lower intensity limiting damage in e.g. imaging applications \cite{Sal98,Per98,Ros09,Ros091,Day04,Lee06,Guz10,Sch12,Sch13}.  Second, time-and-frequency entanglement often allows to obtain higher temporal and spectral resolutions since the two are not Fourier conjugates. Namely, the temporal resolution $\Delta t$ depends on the length of the nonlinear crystal, that is, the entanglement time $T$, while spectral resolution $\Delta\omega$ is determined by the pump envelope. These are independent control variables, not Fourier conjugates and not bound by the uncertainty $\Delta\omega\Delta t\ll1$. We show that entangled photons allow to observe narrow spectral features even in the limit of fast dephasing where the classical line shapes are broad. Elaborate pulse shaping techniques that involve standard prisms compressors and spatial light modulators \cite{Bel03, Pee05, car06, Zah08} can be used to control the amplitude and phase modulation of entangled photon pairs necessary for creating the desired pulse sequence. This can be done using e.g. the Franson interferometer with variable phases and delays in both arms of the interferometer as proposed in \cite{Ray13}. The beam splitters in two arms allow to create four pulses using a single entangled photon pair . In the following we do not specify the experimental details of shaping the pulses, rather we assume a generic sequence of shaped entangled photons.

\section{The loop delay scanning protocol (LOP)}

We consider a model system of an aggregate described by the Frenkel exciton Hamiltonian
\begin{align}
H=H_0+H',
\end{align}
\begin{align}\label{eq:H0}
H_0=\hbar\sum_m\epsilon_mB_m^{\dagger}B_m+\hbar\sum_{m\neq n}J_{mn}B_m^{\dagger}B_n+\hbar\sum_m\frac{\Delta_m}{2}B_m^{\dagger}B_m^{\dagger}B_mB_m,
\end{align}
\begin{align}
H'=E(t)V^{\dagger}+E^{\dagger}(t)V, \quad V^{\dagger}=\sum_mV_m^{\dagger}B_m^{\dagger},
\end{align}
where $H_0$ is the excitonic part, $\epsilon_m$ are site energies, $J_{mn}$ are hopping and $\Delta_m$ is an onsite repulsion (Hubbard type),  and  $B_m$ is an exciton Pauli annihilation operator at site $m$ (e.g. pigment or quantum dot). $H'$ is the dipole interaction with the optical field $E$ in the rotating wave approximation. $E$ is the electric field operator. The eigenstate of Eq. \ref{eq:H0} form distinct exciton bands (see Fig. 1d). In the diagonal eigenstate representation the Hamiltonian for the lowest three manifold of states which are relevant for the present study - ground $g$, single excited $e$ and double excited $f$ manifolds (see Fig 1d) reads
\begin{align}\label{eq:H00}
H_0=\hbar\omega_g|g\rangle\langle g|+\hbar\sum_e\omega_e|e\rangle\langle e|+\hbar\sum_f\omega_f|f\rangle\langle f|,
\end{align}
\begin{align}
V^{\dagger}=\sum_eV_{ge}^{*}|e\rangle\langle g|+\sum_{e,f}V_{ef}^{*}|f\rangle\langle e|.
\end{align}
We consider the following experiment: a sequence of four pulses centered at times $T_a$, $T_b$, $T_c$, and $T_d$ with phases $\phi_a$, $\phi_b$, $\phi_c$, and $\phi_d$ \cite{Pes09}  brings the molecule into its doubly-excite state \cite{Xu10} (see Fig. 1a,b) and  the population of $f$ states is detected $E=\sum_{\alpha=a,b,c,d}E_\alpha e^{i\phi_\alpha}$. This can be done by fluorescence $f\to e$ or after a rapid  internal conversion process the molecule can be deexcited from $f$ to $e$ and fluorescence from $e$ to $g$ is then detected.  We assume that the $e\to g$ and $f\to e$ channels can be distinguished in time or frequency and therefore we can isolate the TPA contributions. Thus, we define the signal as the sum of populations of states $f$.
\begin{align}\label{eq:rff}
S(\Gamma)=\sum_f\rho_{ff}(\Gamma),
\end{align} 
where  $\Gamma$ represents collectively the set of parameters of the incoming pulses. These depend on the protocol and will be specified later.

 \begin{figure}[t]
\begin{center}
\includegraphics[trim=0cm 0cm 0cm 0cm,angle=0, width=0.8\textwidth]{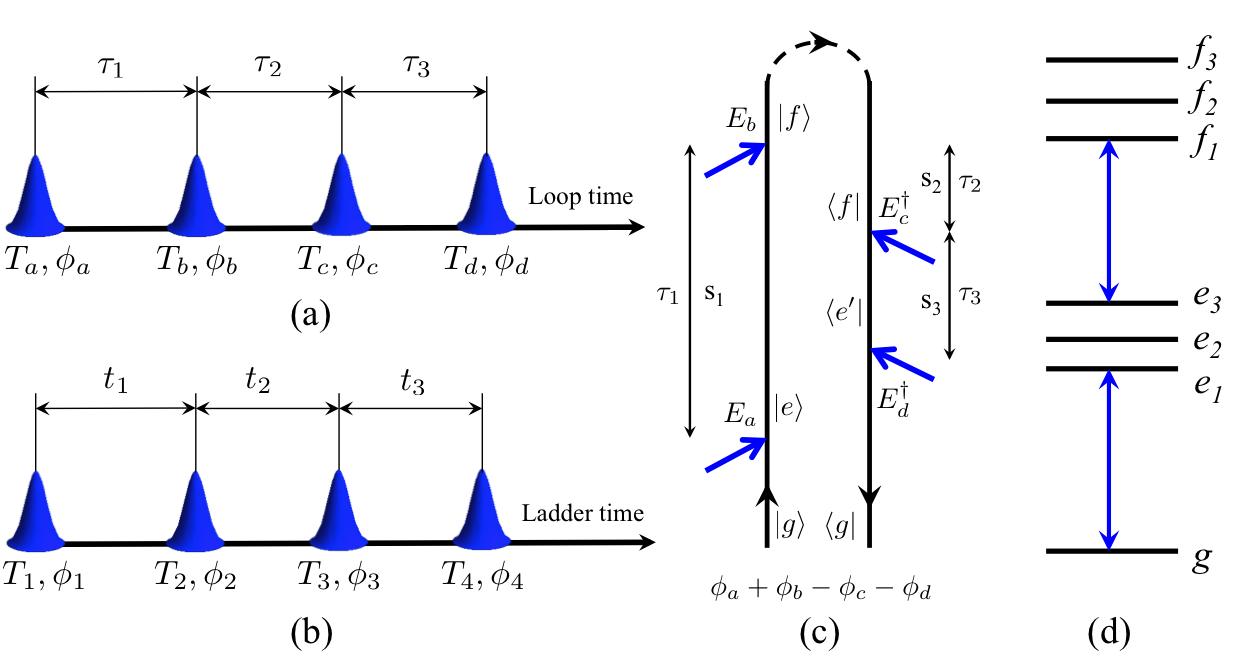}
\end{center}
\caption{(Color online) The pulse sequence for unrestricted LOP \cite{Rah10} - (a), LAP - (b). Loop diagrams for the TPA process with indicated loop delays for the phase cycling selected the signal with $e^{i(\phi_a+\phi_b-\phi_c-\phi_d)}$ - (c). The loop delay variables $s_j$ are centered around $|\tau_j|$, $j=1,2,3$. $s_1$, $s_2$, $s_3$, $\tau_1$, and $\tau_3$ are always positive, $\tau_2$ can be either positive or negative depending on whether the chronologically last interaction occurs with $c$ or $b$. $t_j$, $j=1,2,3$ are always positive. Level scheme for the molecular trimer used in our simulations - (d) (for parameters see Section 5).}
\label{fig:PP}
\end{figure}

The signal (\ref{eq:rff}) for our model is given by the single unrestricted loop diagram in Fig. 1c (for diagram rules see \cite{Rah10}). $a,b,c,d$ denote the pulse sequence {\it ordered along the loop} (not in real time); $a$ represents ``first''.  on the loop etc. Pulses chronologically-ordered in real time will be denoted $1,2,3,4$ which are permutations of $a,b,c,d$ determined by the time arguments, as will be shown below. One can scan various delays $T_\alpha-T_\beta$, $\alpha,\beta=a,b,c,d$ and control the phases $\pm\phi_a\pm\phi_b\pm\phi_c\pm\phi_d$. Phase cycling techniques have been successfully demonstrated as a control tool for the selection of fixed-phase components of  optical signals generated by multiwave mixing \cite{Keu99,Tia03,Tan08,Zha12,Zha121}. Phase cycling can be easily implemented using a pulse shaper by varying the relative inter-pulse phases, which is cycled over 2¹ radians in a number of equally spaced steps \cite{Keu99,Tia03}. To realize the LOP experimentally the indices $a$, $b$, $c$, $d$ are assigned as follows: first by phase cycling we select a signal with phase $\phi_a+\phi_b-\phi_c-\phi_d$. The two pulses with positive phase detection are thus denoted $a$, $b$ and with negative phase - $c$, $d$. In the $a$, $b$ pair pulse $a$ comes first. In the $c$, $d$ pair pulse $d$ comes first. The time variables in Fig. 1c are $\tau_1=T_b-T_a$, $\tau_2=T_c-T_b$, $\tau_3=T_c-T_d$. With this choice $\tau_1$ and $\tau_3$ are positive whereas $\tau_2$ can be either positive or negative. This completely defines the LOP experimentally.

\subsection{Pure states and the loop representation}

In Fig. 1c two interactions with bra- and two - with ket- promote the system to the state described by a population density matrix element $\rho_{ff}$. In the following we omit the phase factor  $e^{i(\phi_a+\phi_b-\phi_c-\phi_d)}$, keeping in mind that all the signals contain it.  The  corresponding signal  (\ref{eq:rff}) can be read-off the diagrams 
\begin{align}\label{eq:S3}
S(\Gamma)=\frac{1}{\hbar^4}\int_{-\infty}^{\infty}dr_a\int_{-\infty}^{\infty}dr_b\int_{-\infty}^{\infty}dr_c\int_{-\infty}^{\infty}dr_d&\langle E_{d}^{\dagger}(r_d)E_{c}^{\dagger}(r_c)E_{b}(r_b)E_{a}(r_a)\rangle\notag\\
\times&\langle\mathcal{T}V(r_d)Vr_c)V^{\dagger}(r_b)V^{\dagger}(r_a)\rangle.
\end{align}
Here $r_\alpha$, $\alpha=a,b,c,d$ are the interaction times of our four pulses with the aggregate, $\mathcal{T}$ denotes the time ordering operator along the loop \cite{Har08}, $G(t)=-i\theta(t)e^{-iHt/\hbar}$ is the Hilbert space Green's function, $\theta(t)$ is the Heaviside step function, and $\mu_{jj'}^{\alpha}=V_{jj'}\cdot\sigma_\alpha$ is the projection of the transition dipole moment $V_{jj'}$, $j,j'=g,e,e',f$ onto the polarization vector $\sigma_\alpha$ of the corresponding field $\alpha=a,b,c,d$. Eq. (\ref{eq:S3}) can be recast using the loop intervals Fig. 1c $s_j$, $j=1,2,3$
\begin{align}\label{eq:S31}
S(\Gamma)=\mathcal{R}\frac{1}{\hbar^4}\int_{-\infty}^{\infty}dr_b\int_0^{\infty}ds_1\int_0^{\infty}ds_2\int_0^{\infty}ds_3&\langle E_{d}^{\dagger}(r_b-s_2-s_3)E_{c}^{\dagger}(r_b-s_2)E_{b}(r_b)E_{a}(r_b-s_1)\rangle\notag\\\times&\langle V(r_b-s_2-s_3)V(r_b-s_2)V^{\dagger}(r_b)V^{\dagger}(r_b-s_1).
\end{align}
Time ordering is now explicitly specified by the integration limits and we no longer need the time ordering operator. In this expression $s_2$ is positive (interaction with pulse $b$ is chronologically the last). The contribution where the field $c$ is the last  is included by taking the real part $\mathcal{R}$. 

One can alternatively recast Eq. (\ref{eq:S31}) in frequency-domain using the electric field operators $\mathcal{E}_\alpha(t)=\int_{-\infty}^{\infty}\frac{d\omega}{2\pi}E(\omega)e^{-i\omega(t-T_\alpha)}$, $\alpha=a,b,c,d$
\begin{align}\label{eq:S4}
S_{LOP}(\tau_1,\tau_2,&\tau_3)=\mathcal{R}\frac{i}{\hbar^4}\int_{-\infty}^{\infty}\frac{d\omega_a}{2\pi}\int_{-\infty}^{\infty}\frac{d\omega_b}{2\pi}\int_{-\infty}^{\infty}\frac{d\omega_d}{2\pi} \langle E^{\dagger}(\omega_d)E^{\dagger}(\omega_a+\omega_b-\omega_d)E(\omega_b)E(\omega_a)\rangle \notag\\
&\times \sum_{e,e',f}\mu_{ge'}^a\mu_{e'f}^b\mu_{fe}^{c*}\mu_{eg}^{d*}G_{e'}^{\dagger}(\omega_d)G_e(\omega_a)[G_f(\omega_a+\omega_b)\theta(\tau_2)-G_f^{\dagger}(\omega_a+\omega_b)\theta(-\tau_2)]\notag\\
&\times e^{-i\omega_a\tau_1+i\omega_d\tau_3-i(\omega_a+\omega_b)\tau_2}\theta(\tau_1)\theta(\tau_3),
\end{align} 
where the LOP control variables $\tau_1=T_b-T_a$, $\tau_2=T_c-T_b$, $\tau_3=T_c-T_d$ are the delays between pulse centers and $G(\omega)=1/[\omega+\omega_g-H/\hbar+i\epsilon]$ is a frequency domain Green's function. 

In the frequency-domain the field correlation function is defined as a Fourier transform of the time-domain field correlation function
\begin{align}\label{eq:Et}
&\langle E^{\dagger}(\omega_d)E^{\dagger}(\omega_c)E(\omega_b)E(\omega_a)\rangle\notag\\
&=\int_{-\infty}^{\infty}dt_1'\int_{-\infty}^{\infty}dt_2'\int_{-\infty}^{\infty}dt_3'\int_{-\infty}^{\infty}dt_4'e^{i\omega_at_1'+i\omega_bt_2'-i\omega_ct_3'-i\omega_dt_4'}\langle E_{d}^{\dagger}(t_3')E_{c}^{\dagger}(t_4')E_{b}(t_2')E_{a}(t_1')\rangle.
\end{align} 
In Eq. (\ref{eq:S4}) we used Eq. (\ref{eq:Et}) and the time translation invariance symmetry which implies $\omega_a+\omega_b-\omega_c-\omega_d=0$.

 In the absence of a bath, the matter correlation function is given by
\begin{align}\label{eq:Vt}
&\sum_{e,e',f}\mu_{ge'}^a\mu_{e'f}^b\mu_{fe}^{c*}\mu_{eg}^{d*}G_{e'}^{\dagger}(\omega_d)G_e(\omega_a)[G_f(\omega_a+\omega_b)\theta(\tau_2)-G_f^{\dagger}(\omega_a+\omega_b)\theta(-\tau_2)]\notag\\
&=\int_{-\infty}^{\infty}dt_1'\int_{-\infty}^{\infty}dt_2'\int_{-\infty}^{\infty}dt_3'\int_{-\infty}^{\infty}dt_4'e^{-i\omega_at_1'-i\omega_bt_2'+i\omega_ct_3'+i\omega_dt_4'}\langle\mathcal{T}V_{ge'}(t_3')V_{e'f}(t_4')V_{fe}^{\dagger}(t_2')V_{eg}^{\dagger}(t_1')\rangle
\end{align}


 
 A Fourier transform of (\ref{eq:S4}) with respect to loop delays then gives a 3D signal
\begin{align}\label{eq:S5}
S_{LOP}(\Omega_1,\Omega_2,\Omega_3)&=\int_{-\infty}^{\infty}d\tau_1\int_{-\infty}^{\infty}d\tau_2\int_{-\infty}^{\infty}d\tau_3e^{i\Omega_1\tau_1+i\Omega_2\tau_2+i\Omega_3\tau_3}S_{LOP}(\tau_1,\tau_2,\tau_3).
\end{align}
Combining Eqs. (\ref{eq:S3}) - (\ref{eq:S5}) gives
\begin{align}
S_{LOP}(\Omega_1,\Omega_2,\Omega_3)=S'_{LOP}(\Omega_1,\Omega_2,\Omega_3)+S_{LOP}^{'*}(-\Omega_1,-\Omega_2,-\Omega_3),
\end{align}
where
\begin{align}\label{eq:S57}
S'_{LOP}&(\Omega_1,\Omega_2,\Omega_3)=\frac{2}{\hbar^4}\int_{-\infty}^{\infty}\frac{d\omega_a}{2\pi}\int_{-\infty}^{\infty}\frac{d\omega_d}{2\pi}\langle E_{d}^{\dagger}(\omega_d)E_{c}^{\dagger}(\omega_{fg}-\omega_d)E_{b}(\omega_{fg}-\omega_a)E_{a}(\omega_a)\rangle \notag\\
&\times\sum_{e,e',f} \frac{\mu_{ge'}^a\mu_{e'f}^b\mu_{fe}^{c*}\mu_{eg}^{d*}G_e(\omega_a)G_{e'}^{\dagger}(\omega_d)}{[\omega_a-\Omega_1-i\epsilon][\omega_d+\Omega_3+i\epsilon]}\left[\frac{G_f(\omega_a+\omega_b)}{\omega_a+\omega_b-\Omega_2-i\epsilon}+\frac{G_f^{\dagger}(\omega_a+\omega_b)}{\omega_a+\omega_b-\Omega_2+i\epsilon}\right].
\end{align}
where the limit $\epsilon\to 0 $ is understood.  One can then evaluate the remaining frequency integrals in Eq. (\ref{eq:S57}) for a given light field correlation function using residue calculus.

So far we did not specify the nature of the field, and Eqs. (\ref{eq:S4}) - (\ref{eq:S57}) hold for arbitrary type of field, be it classical, stochastic or entangled. All relevant field information is contained in its four point field correlation function which must be evaluated separately. For classical coherent fields this function factorizes (in time or frequency) into a product of four amplitudes. Otherwise for entangled or stochastic fields the correlation function causes a coupling between two interaction times, which affects the signals.

\subsection{Pure dephasing, bath effects and the ladder representation}

 \begin{figure}[t]
\begin{center}
\includegraphics[trim=0cm 0cm 0cm 0cm,angle=0, width=0.8\textwidth]{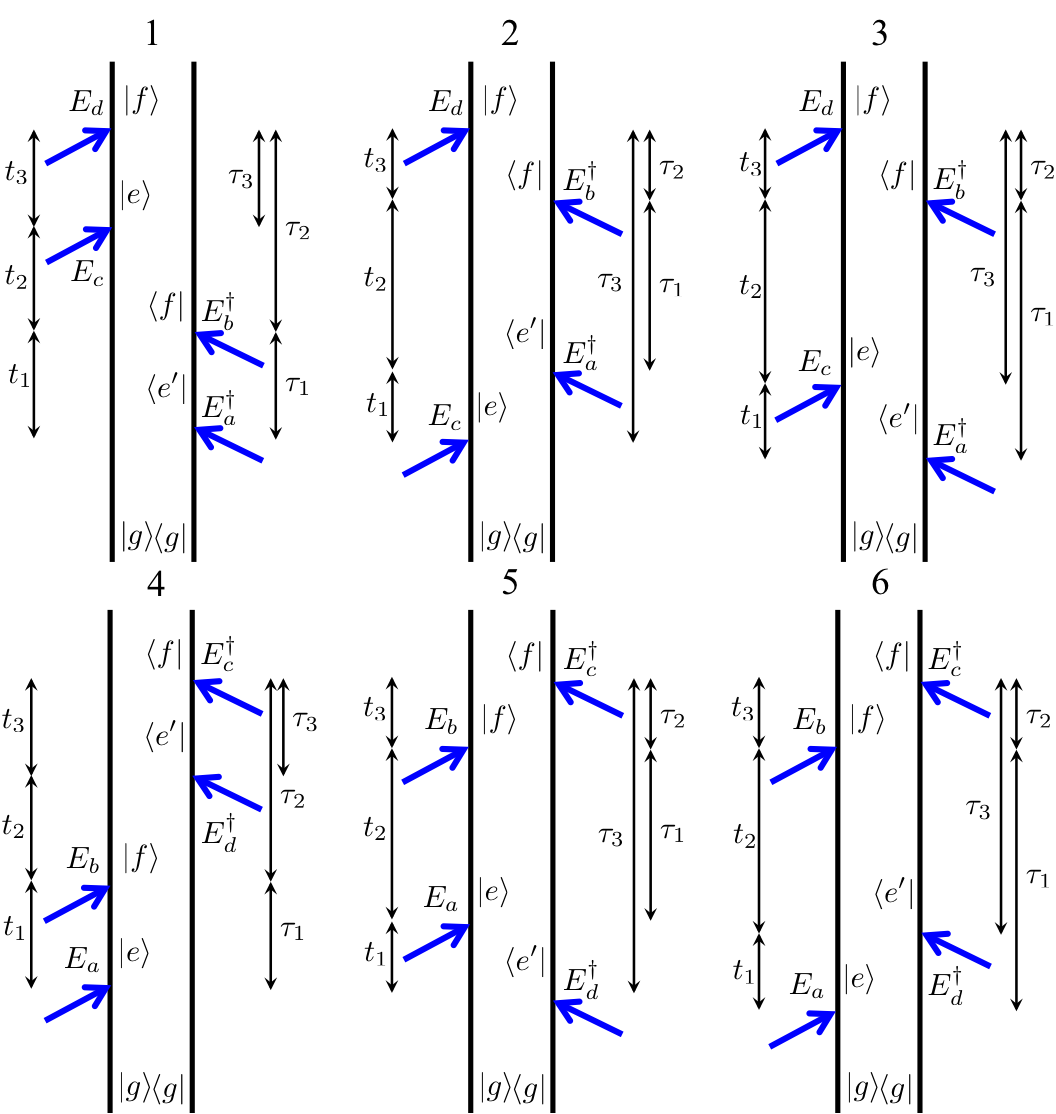}
\end{center}
\caption{(Color online) Ladder diagrams for the TPA signal with selected phase $e^{i(\phi_a+\phi_b-\phi_c-\phi_d)}$. Both loop $\tau_j$ and ladder $t_j$ delays, $j=1,2,3$ are indicated. The transformation between two is different for each diagram. Time translation invariance implies $\omega+\omega_b-\omega_c-\omega_d=0$. The LOP signal is a sum of all six diagrams whereas the LAP  can be separated into $k_I$, $k_{II}$ and $k_{III}$ signals (see text).}
\label{fig:PP}
\end{figure}

When the exciton system is coupled to a bath, it can no longer be described by a wavefunction once the bath is eliminated. To evaluate the loop diagram it must be broken into several ladder diagrams (for notation see \cite{Rah10}) which represent the density matrix. The unrestricted loop diagram in Fig. 1b is split into the six ladder diagrams shown in Fig. 1c and the signal (\ref{eq:S4}) is given by  sum of all six terms $S_{LOP}(\tau_1,\tau_2,\tau_3)=\sum_{j=1}^{6}S_{LOP}^{(j)}(\tau_1,\tau_2,\tau_3)$ where
\begin{align}\label{eq:S60}
S_{LOP}^{(j)}&(\tau_1,\tau_2,\tau_3)=\int_{-\infty}^{\infty}\frac{d\omega_a}{2\pi}\frac{d\omega_b}{2\pi}\frac{d\omega_d}{2\pi}D_{LOP}^{(j)}(\tau_1,\tau_2,\tau_3;\omega_a,\omega_b,\omega_d)\tilde{S}^{(j)}(\omega_a,\omega_b,\omega_d)-c.c.,
\end{align}
where 
\begin{align}\label{eq:Dlop1}
D_{LOP}^{(j)}(\tau_1,\tau_2,\tau_3;\omega_a,\omega_b,\omega_d)=\theta(\tau_1)\theta_j(\pm\tau_2)\theta(\tau_3)e^{-i\omega_a\tau_1+i\omega_d\tau_3-i(\omega_a+\omega_b)\tau_2},
\end{align}
is a display function which depends on the control parameters specific to the chosen protocol. In $\theta_j(\pm\tau_2)$ the ``minus'' sign applies for diagrams $j=1,2,3$ and the ``plus'' sign for $j=4,5,6$,
\begin{align}\label{eq:rffj}
\tilde{S}^{(j)}(\omega_a,\omega_b,\omega_d)&=\langle E^{\dagger}(\omega_d)E^{\dagger}(\omega_a+\omega_b-\omega_d)E(\omega_b)E(\omega_a)\rangle R^{(j)}(\omega_a,\omega_b,\omega_d),
\end{align}
and
\begin{align}\label{eq:S61}
&R^{(1)}(\omega_a,\omega_b,\omega_d)=\hbar^{-4}\sum_{e,e',f}\mu_{ge'}^a\mu_{e'f}^b\mu_{fe}^{c*}\mu_{eg}^{d*}\mathcal{G}_{ge'}(-\omega_d)\mathcal{G}_{ef}(-\omega_b)\mathcal{G}_{gf}(-\omega_a-\omega_b),\notag\\
&R^{(2)}(\omega_a,\omega_b,\omega_d)= \hbar^{-4}\sum_{e,e',f}\mu_{ge'}^a\mu_{e'f}^b\mu_{fe}^{c*}\mu_{eg}^{d*}\mathcal{G}_{eg}(\omega_a)\mathcal{G}_{ef}(-\omega_b)\mathcal{G}_{ee'}(\omega_a-\omega_d),\notag\\
&R^{(3)}(\omega_a,\omega_b,\omega_d)= \hbar^{-4}\sum_{e,e',f}\mu_{ge'}^a\mu_{e'f}^b\mu_{fe}^{c*}\mu_{eg}^{d*}\mathcal{G}_{ge'}(-\omega_d)\mathcal{G}_{ef}(-\omega_b)\mathcal{G}_{ee'}(\omega_a-\omega_d),\notag\\
&R^{(4)}(\omega_a,\omega_b,\omega_d)= \hbar^{-4}\sum_{e,e',f}\mu_{ge'}^a\mu_{e'f}^b\mu_{fe}^{c*}\mu_{eg}^{d*}\mathcal{G}_{eg}(\omega_a)\mathcal{G}_{fe'}(\omega_a+\omega_b-\omega_d)\mathcal{G}_{fg}(\omega_a+\omega_b),\notag\\
&R^{(5)}(\omega_a,\omega_b,\omega_d)= \hbar^{-4}\sum_{e,e',f}\mu_{ge'}^a\mu_{e'f}^b\mu_{fe}^{c*}\mu_{eg}^{d*}\mathcal{G}_{ge'}(-\omega_d)\mathcal{G}_{fe'}(\omega_a+\omega_b-\omega_d)\mathcal{G}_{ee'}(\omega_a-\omega_d),\notag\\
&R^{(6)}(\omega_a,\omega_b,\omega_d)=\hbar^{-4}\sum_{e,e',f}\mu_{ge'}^a\mu_{e'f}^b\mu_{fe}^{c*}\mu_{eg}^{d*}\mathcal{G}_{eg}(\omega_a)\mathcal{G}_{fe'}(\omega_a+\omega_b-\omega_d)\mathcal{G}_{ee'}(\omega_a-\omega_d).
\end{align}
Here we had introduced the Liouville space Green's function 
\begin{align}
\mathcal{G}_{\alpha\beta}(\omega)=-i\int_0^{\infty}dte^{i\omega t}\langle G_\alpha(t)G_\beta^{\dagger}(t)\rangle_B,
\end{align}
 where $\langle...\rangle_B=\text{Tr}[...\rho_B]$ represents the trace over the bath degrees of freedom. The bra and the ket evolutions (and the corresponding time variables) are now coupled by the bath.   The effect of couplings between interaction times due to nonclassical field  is by evaluating the frequency integrals in Eq. (\ref{eq:S60}) using time-domain display function in Eq. (\ref{eq:Dlop1}). The result for entangled photons is given in Appendix A.
 To see the effect on the mixing of the frequency variables we then take a Fourier transform of Eq. (\ref{eq:S60}) with respect to loop delay variable $\tau_j$, $j=1,2,3$ and obtain the signal
  \begin{align}\label{eq:S70}
S_{LOP}(\Omega_1,\Omega_2,\Omega_3)=\sum_{j=1}^6\int_{-\infty}^{\infty}\frac{d\omega_a}{2\pi}&\frac{d\omega_b}{2\pi}\frac{d\omega_d}{2\pi}[D_{LOP}^{(j)}(\Omega_1,\Omega_2,\Omega_3;\omega_a,\omega_b,\omega_d)\tilde{S}^{(j)}(\omega_a,\omega_b,\omega_d)\notag\\
&-D_{LOP}^{(j)*}(-\Omega_1,-\Omega_2,-\Omega_3;\omega_a,\omega_b,\omega_d)\tilde{S}^{(j)*}(\omega_a,\omega_b,\omega_d)],
\end{align}
where
\begin{align}\label{eq:Fjlp}
D_{LOP}^{(j)}(\Omega_1,\Omega_2,\Omega_3;\omega_a,\omega_b,\omega_d)=\frac{\mp i}{[\omega_a-\Omega_1-i\epsilon][\Omega_3+\omega_d+i\epsilon][\Omega_2-\omega_a-\omega_b\mp i\epsilon]},
\end{align}
and minus (plus) sign corresponds to contributions of diagrams $1-3$ ($4-6$). The coupling between interaction times now translates into a mixing of their conjugate frequency variables $\Omega_j$, $j=1,2,3$. The 3D signals (\ref{eq:S70}) are given by a 3D spectral overlap between Green's functions of the matter and field, where the latter are governed by $[\Omega_j-\omega_\alpha\pm i\epsilon]^{-1}$ dressed by a four point field correlation function which selects the field-matter pathways. The response of the system to classical light fields is given by nonlinear response functions which can be expressed by sums over various quantum pathways of matter. In the case of quantum field the response is typically treated in the joint field-matter space to account for back-reaction and other nonclassical effects of the field. In this case the response is summed over various quantum pathways in the joint field-matter space. Depending on the field parameters some quantum pathways can be suppressed or enhanced. The field correlation function controls the relevant spectral range of the pathways that contribute to the signal. Different integrations may couple various frequencies $\omega_\alpha$, $\alpha=a,b,d$ into a single field-matter Green's function. Upon evaluating the relevant frequency integrations different $\Omega_j$, $j=1,2,3$ will be coupled. This will result in various cross-peaks between $\Omega_j$ variables, as becomes apparent by comparing a field contribution in Eq. (\ref{eq:Fjlp}) with various responses in Eq. (\ref{eq:S61}). Together with the bath dephasing effects, the relevant spectral width of these cross-peaks can vary significantly compared to that of the system without bath interacting with classical fields. Below we will investigate the signatures of the bath and the state of field in the signals.

\section{The ladder delay scanning protocol (LAP)}

In standard multidimensional techniques the time variables represent the pulses as they interact with sample in chronological order \cite{Muk95}. These are conveniently given by the ladder delays. In the LOP the time ordering between pulses is maintained only on each branch of the loop but not between branches. The LAP in contrast involves full time-ordering of all four pulses.  The arrival time of the various pulses in chronological order is $T_1<T_2<T_3<T_4$. The indices $1,2,3,4$ are some permutation of  $a,b,c,d$ depending on the diagram. The ladder delays are defined as $t_1=T_2-T_1$, $t_2=T_3-T_2$, $t_3=T_4-T_3$. Ladder diagrams keep track of chronological delays. Each ladder diagram will have its own set of relations between $t_j$, $j=1,2,3$ and pulse delays $T_\alpha-T_\beta$, $\alpha,\beta=a,b,c,d$. One can then use the phase cycling to select the diagrams shown in Fig. 2 e.g. $\mathbf{k}_I=-\mathbf{k}_1+\mathbf{k}_2+\mathbf{k}_3$, $\mathbf{k}_{II}=\mathbf{k}_1-\mathbf{k}_2+\mathbf{k}_3$ and $\mathbf{k}_{III}=\mathbf{k}_1+\mathbf{k}_2-\mathbf{k}_3$. This gives 
\begin{align}\label{eq:k1}
S_{\mathbf{k}_I}(t_1,t_2,t_3)=S_{LAP}^{(2)}(t_1,t_2,t_3)+S_{LAP}^{(5)}(t_1,t_2,t_3),
\end{align} 
\begin{align}\label{eq:k2}
S_{\mathbf{k}_{II}}(t_1,t_2,t_3)=S_{LAP}^{(3)}(t_1,t_2,t_3)+S_{LAP}^{(6)}(t_1,t_2,t_3),
\end{align} 
\begin{align}\label{eq:k3}
S_{\mathbf{k}_{III}}(t_1,t_2,t_3)=S_{LAP}^{(1)}(t_1,t_2,t_3)+S_{LAP}^{(4)}(t_1,t_2,t_3).
\end{align} 
where
\begin{align}\label{eq:S80}
S_{LAP}^{(j)}(t_1,t_2,t_3)=&\int_{-\infty}^{\infty}\frac{d\omega_a}{2\pi}\frac{d\omega_b}{2\pi}\frac{d\omega_d}{2\pi}D_{LAP}^{(j)}(t_1,t_2,t_3;\omega_a,\omega_b,\omega_d)\tilde{S}^{(j)}(\omega_a,\omega_b,\omega_d)-c.c..
\end{align}
Here the LAP display functions are given by
\begin{align}\label{eq:S81}
D_{LAP}^{(1)}(t_1,t_2,t_3;\omega_a,\omega_b,\omega_d)=\theta(t_1)\theta(t_2)\theta(t_3)e^{i\omega_bt_3+i\omega_dt_1+i(\omega_a+\omega_b)t_2},
\end{align}
where $t_3=T_b-T_a$, $t_2=T_a-T_c$, $t_1=T_c-T_d$.
\begin{align}\label{eq:S82}
D_{LAP}^{(2)}(t_1,t_2,t_3;\omega_a,\omega_b,\omega_d)=\theta(t_1)\theta(t_2)\theta(t_3)e^{i\omega_bt_3-i\omega_at_1+i(\omega_d-\omega_a)t_2},
\end{align}
where $t_3=T_b-T_c$, $t_2=T_c-T_d$, $t_1=T_d-T_a$.
\begin{align}\label{eq:S83}
D_{LAP}^{(3)}(t_1,t_2,t_3;\omega_a,\omega_b,\omega_d)=\theta(t_1)\theta(t_2)\theta(t_3)e^{i\omega_bt_3+i\omega_dt_1+i(\omega_d-\omega_a)t_2},
\end{align}
where $t_3=T_b-T_c$, $t_2=T_c-T_a$, $t_1=T_a-T_d$.
\begin{align}\label{eq:S84}
D_{LAP}^{(4)}(t_1,t_2,t_3;\omega_a,\omega_b,\omega_d)=\theta(t_1)\theta(t_2)\theta(t_3)e^{-i(\omega_a+\omega_b-\omega_d)t_3-i\omega_at_1-i(\omega_a+\omega_b)t_2},
\end{align}
where $t_3=T_c-T_d$, $t_2=T_d-T_b$, $t_1=T_b-T_a$.
\begin{align}\label{eq:S85}
D_{LAP}^{(5)}(t_1,t_2,t_3;\omega_a,\omega_b,\omega_d)=\theta(t_1)\theta(t_2)\theta(t_3)e^{-i(\omega_a+\omega_b-\omega_d)t_3+i\omega_dt_1+i(\omega_d-\omega_a)t_2},
\end{align}
where $t_3=T_c-T_b$, $t_2=T_b-T_a$, $t_1=T_a-T_d$.
\begin{align}\label{eq:S86}
D_{LAP}^{(6)}(t_1,t_2,t_3;\omega_a,\omega_b,\omega_d)=\theta(t_1)\theta(t_2)\theta(t_3)e^{-i(\omega_a+\omega_b-\omega_d)t_3-i\omega_at_1+i(\omega_d-\omega_a)t_2},
\end{align}
where $t_3=T_c-T_b$, $t_2=T_b-T_d$, $t_1=T_d-T_a$. The corresponding expressions for entangled photons are given in Appendix B. 

We now take the Fourier transform with respect to ladder delay variable $t_j$, $j=1,2,3$
  \begin{align}\label{eq:S90}
S_{LAP}^{(j)}(\tilde{\Omega}_1,\tilde{\Omega}_2,\tilde{\Omega}_3)=\int_{-\infty}^{\infty}dt_1\int_{-\infty}^{\infty}dt_2\int_{-\infty}^{\infty}d&t_3e^{i\tilde{\Omega}_1t_1+i\tilde{\Omega}_2t_2+i\tilde{\Omega}_3t_3}S_{LAP}^{(j)}(t_1,t_2,t_3)\end{align}
This gives
\begin{align}
S_{LAP}^{(j)}(\tilde{\Omega}_1,\tilde{\Omega}_2,\tilde{\Omega}_3)=\int_{-\infty}^{\infty}\frac{d\omega_a}{2\pi}\frac{d\omega_b}{2\pi}&\frac{d\omega_d}{2\pi}[D_{LAP}^{(j)}(\tilde{\Omega}_1,\tilde{\Omega}_2,\tilde{\Omega}_3;\omega_a,\omega_b,\omega_d)\tilde{S}^{(j)}(\omega_a,\omega_b,\omega_d)\notag\\
&-D_{LAP}^{(j)*}(-\tilde{\Omega}_1,-\tilde{\Omega}_2,-\tilde{\Omega}_3;\omega_a,\omega_b,\omega_d)\tilde{S}^{(j)*}(\omega_a,\omega_b,\omega_d)],
\end{align}
where
\begin{align}\label{eq:Fjld}
&D_{LAP}^{(1)}(\Omega_1,\Omega_2,\Omega_3,\omega_a,\omega_b,\omega_d)=\frac{1}{[\Omega_3+\omega_b+i\epsilon][\Omega_1+\omega_d+i\epsilon][\Omega_2+\omega_a+\omega_b+i\epsilon]},\notag\\
&D_{LAP}^{(2)}(\Omega_1,\Omega_2,\Omega_3,\omega_a,\omega_b,\omega_d)=\frac{1}{[\Omega_3+\omega_b+i\epsilon][\Omega_1-\omega_a+i\epsilon][\Omega_2+\omega_d-\omega_a+i\epsilon]},\notag\\
&D_{LAP}^{(3)}(\Omega_1,\Omega_2,\Omega_3,\omega_a,\omega_b,\omega_d)=\frac{1}{[\Omega_3+\omega_b+i\epsilon][\Omega_1+\omega_d+i\epsilon][\Omega_2+\omega_d-\omega_a+i\epsilon]},\notag\\
&D_{LAP}^{(4)}(\Omega_1,\Omega_2,\Omega_3,\omega_a,\omega_b,\omega_d)=\frac{1}{[\Omega_3-\omega_a-\omega_b+\omega_d+i\epsilon][\Omega_1-\omega_a+i\epsilon][\Omega_2-\omega_a-\omega_b+i\epsilon]},\notag\\
&D_{LAP}^{(5)}(\Omega_1,\Omega_2,\Omega_3,\omega_a,\omega_b,\omega_d)=\frac{1}{[\Omega_3-\omega_a-\omega_b+\omega_d+i\epsilon][\Omega_1+\omega_d+i\epsilon][\Omega_2+\omega_d-\omega_a+i\epsilon]},\notag\\
&D_{LAP}^{(6)}(\Omega_1,\Omega_2,\Omega_3,\omega_a,\omega_b,\omega_d)=\frac{1}{[\Omega_3-\omega_a-\omega_b+\omega_d+i\epsilon][\Omega_1-\omega_a+i\epsilon][\Omega_2+\omega_d-\omega_a+i\epsilon]}.
\end{align}
We note that the frequency variables $\omega_\alpha$, $\alpha=a,b,d$  in Eq. (\ref{eq:Fjld}) are the same combinations that appear in the matter responses in Eq. (\ref{eq:S61}). This means that  the signal will factorize into a product of several Green's functions with uncoupled frequency arguments $\tilde{\Omega}_j$, $j=1,2,3$. Of course this holds only in the absence of additional sources of correlating the variables caused by e.g. dephasing (bath) or the state of light. In the LOP, in contrast,  the time correlations that result in the frequency mixing is apparent.  Different frequency components $\omega_\alpha$, $\alpha=a,b,d$ that enter the Green's function in Eq. (\ref{eq:S61}) interfere when convoluted with the same display function Eqs. (\ref{eq:Fjlp}).

\section{Classical vs quantum light fields}

The state of light that enters the signal via the four-point frequency domain correlation function of the electric field in Eq. (\ref{eq:rffj}) can mix various frequency variables which arise from the coupling between the interaction times. In the following we consider the  twin photon entangled state of light and compare it to the classical (coherent) state. Ideal multidimensional techniques use impulsive fields well separated in time with infinite bandwidth. However as shown in the following it is crucial to keep the finite bandwidth.

In the case of classical light, the four-point correlation function simply factorizes into a product of four electric field amplitudes
\begin{align}
\langle E^{\dagger}(\omega_d)E^{\dagger}(\omega_a+\omega_b-\omega_d)E(\omega_b)E(\omega_a)\rangle=\mathcal{E}^{*}(\omega_d)\mathcal{E}^{*}(\omega_a+\omega_b-\omega_d)\mathcal{E}(\omega_b)\mathcal{E}(\omega_a).
\end{align}
 Note, that because classical fields do not impose correlations between various interaction times, either LOP or LAP can be used. In the following simulations we assume lorentzian pulses and set $\mathcal{E}(\omega)=\langle E(\omega)\rangle=A_1/[\omega-\omega_p+i\sigma_p]$. 
 
 Twin photons are created via type-I spontaneous parametric down conversion of a classical pump pulse with frequency $2\omega_p$ into a pair of photons with central frequencies $\omega_1^{(0)}$ and $\omega_2^{(0)}$. For the degenerate process $\omega_1^{(0)}=\omega_2^{(0)}=\omega_p$ the quantum state of light is given by the wave function
\begin{align}
|\psi\rangle=\int_{-\infty}^{\infty}\frac{d\omega_1}{2\pi}\frac{d\omega_2}{2\pi}\Phi(\omega_1,\omega_2)a_{\omega_1}^{\dagger}a_{\omega_2}^{\dagger}|0\rangle,
\end{align}
where $a_\omega^{\dagger}$ is the photon creation operator in the frequency mode $\omega$ and $\Phi(\omega_1,\omega_2)$ is the two-photon amplitude
\begin{align}
\Phi(\omega_1,\omega_2)=A_p(\omega_1+\omega_2)\text{sinc}\left[\frac{\omega_1-\omega_p}{2}T_{1e}+\frac{\omega_2-\omega_p}{2}T_{2e}\right]e^{i\frac{\omega_1-\omega_p}{2}T_{1e}+i\frac{\omega_2-\omega_p}{2}T_{2e}}+(T_{1e}\leftrightarrow T_{2e}),
\end{align}
where $A_p(\omega)=A_0/[\omega-2\omega_p+i\sigma_p]$ is the lorentzian classical pump pulse envelope. The variables $T_{1e}=L/v_p-L/v_1$ and $T_{2e}=L/v_p-L/v_2$ represent the time delays  between the various beams acquired in the course of the propagation through the crystal with length $L$. Here,  $v_p$, $v_{1,2}$ denote the group velocity of the pump pulse, or beams 1 and 2,
respectively. The entanglement time $T_{e}=T_{2e}-T_{1e}$ along with the pump bandwidth $\sigma_p$ are the two key parameters that define the degree of correlation between twin photons. The four-point field correlation function in Eq. (\ref{eq:rffj}) is now given by
\begin{align}
\langle E^{\dagger}(\omega_d)E^{\dagger}(\omega_a+\omega_b-\omega_d)E(\omega_b)E(\omega_a)\rangle=\Phi^{*}(\omega_a+\omega_b-\omega_d,\omega_d)\Phi(\omega_a,\omega_b).
\end{align}
It is important to note that since the four-point correlation function of the entangled twin state factorizes into a product of two two-point correlation functions of the form $\langle E(\omega_b)E(\omega_a)\rangle$ it only couples different interaction times within the bra- ($T_a$, $T_b$) and within the ket- ($T_c$, $T_d$). This means that the coupling between the interaction times in this case occurs on one branch of the loop and interaction times on different branches are not coupled. LOP thus offers a natural scanning protocol for quantum spectroscopy with entangled twin-state of light.


\section{Simulations}

We have simulated the signal (\ref{eq:S70}) using the LOP protocol and compared it with the standard fully time ordered LAP protocol given by Eq. (\ref{eq:S90}) for a model trimer described by the Frenkel exciton Hamiltonian (\ref{eq:H0}). We first present LOP results for classical and entangled light. We then provide the reference by demonstrating the LAP results. The parameters used are $\epsilon_1=1.518$ eV, $\epsilon_2=1.530$ eV, $\epsilon_3=1.526$ eV, $J_{12}=10$ meV, $J_{13}=2$ meV, $J_{13}=3$ meV. All three chromophores have the same transition dipole $V_1=V_2=V_3$. In the eigenstate basis the Hamiltonian (\ref{eq:H00}) has parameters $\hbar\omega_{e_1}=1.512$ eV, $\hbar\omega_{e_2}=1.525$ eV, $\hbar\omega_{e_3}=1.537$ eV, $\hbar\omega_{f_1}=3.044$ eV, $\hbar\omega_{f_2}=3.048$ eV, and $\hbar\omega_{f_3}=3.056$ eV. We focus on the three exciton bands: $g$, $e$, and $f$. In our model we have two sources of dephasing. First intraband dephasing which is associated with transition within excited state band, e.g. $e-eÕ$. Second interband dephasing that governs the transitions between e.g single and double excited states  $e-f$.

 \begin{figure}[t]
\begin{center}
\includegraphics[trim=0cm 0cm 0cm 0cm,angle=0, width=0.8\textwidth]{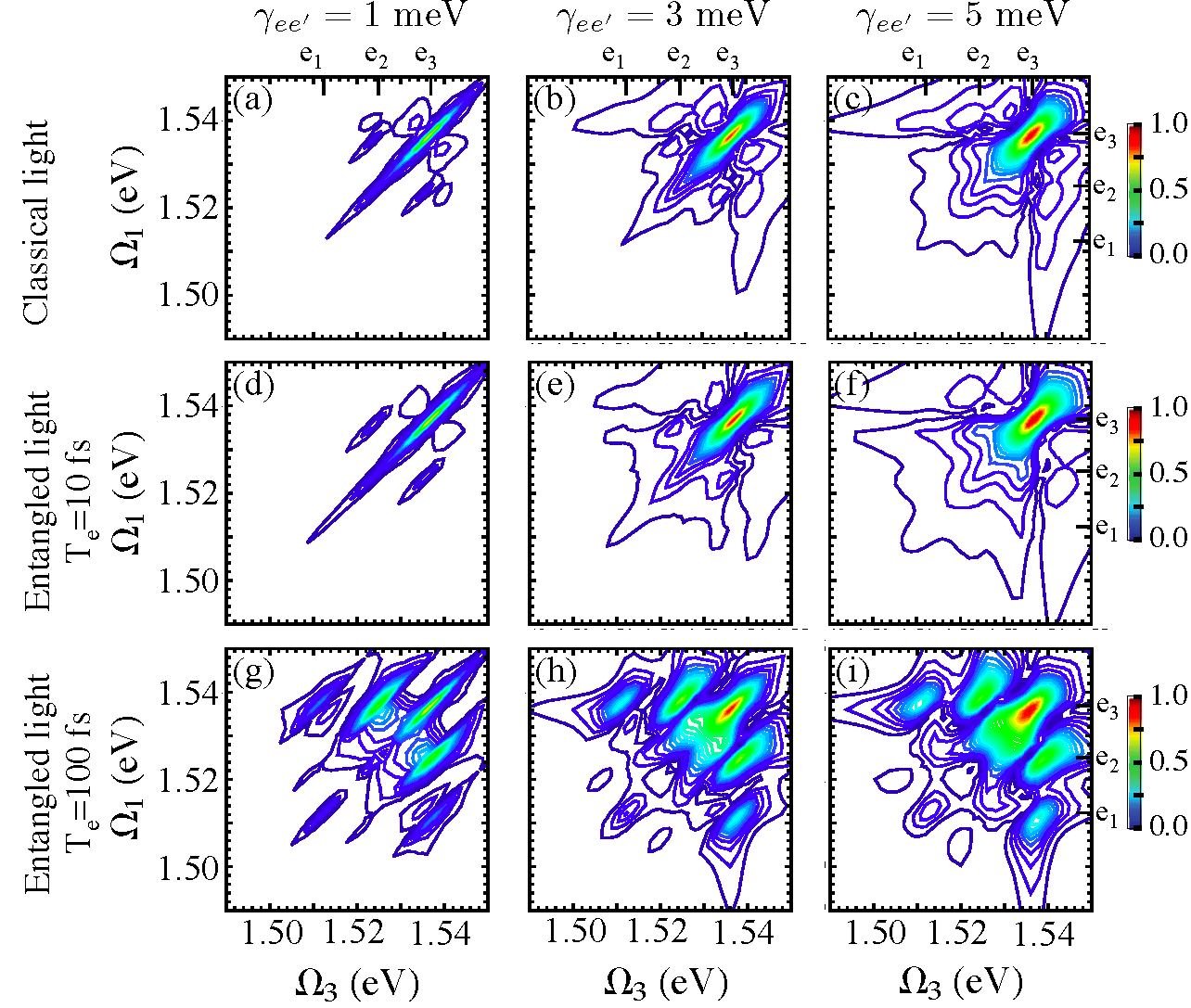}
\end{center}
\caption{(Color online) $S_{LOP}(\Omega_1,\tau_2=0,\Omega_3)$ Eq. (\ref{eq:S70}) for a molecular trimer using classical light - top row, entangled light with $T_e=10$ fs - middle row and $T_e=100$ fs - bottom row. Intraband dephasing $\gamma_{ee'}=1$ meV - left column, $3$ meV - middle column and $5$ meV - right column. $\gamma_{eg}=\gamma_{fe}=\gamma_{fg}=4$ meV, $\sigma_p=20$ meV, $2\omega_p=3.0621$ eV. All other parameters are given in the beginning of Section 5.}
\label{fig:PP}
\end{figure}

\subsection{LOP signals}

 \begin{figure}[t]
\begin{center}
\includegraphics[trim=0cm 0cm 0cm 0cm,angle=0, width=0.8\textwidth]{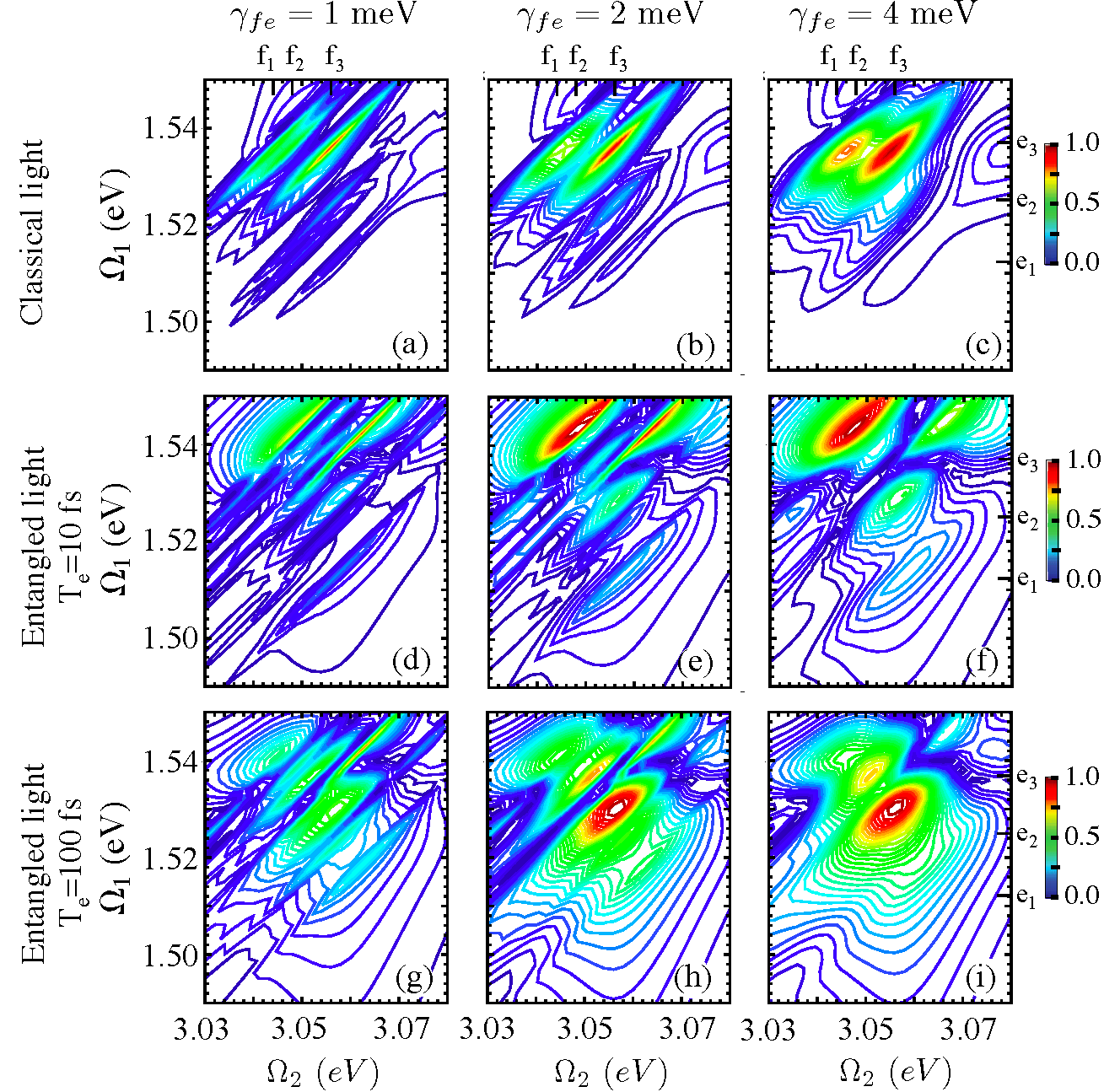}
\end{center}
\caption{(Color online) $S_{LOP}(\Omega_1,\Omega_2,\tau_3=0)$ Eq. (\ref{eq:S70})  for a molecular trimer using classical light - top row, entangled light with $T_e=10$ fs - middle row and $T_e=100$ fs - bottom row. Interband dephasing $\gamma_{fe}=1$ meV - left column, $2$ meV - middle column, and $4$ meV - right column. $\gamma_{eg}=\gamma_{fg}=\gamma_{ee'}=10$ meV. All other parameters are the same as in Fig. 3.}
\label{fig:PP}
\end{figure}

Below we present  two-dimensional signals obtained by setting one time interval to zero. Fig. 3 shows the simulated $S_{LOP}(\Omega_1,\tau_2=0,\Omega_3)$ for a trimer using classical light (top row) and entangled light (mid and bottom row). This signal reveals the intraband dephasing rate $\gamma_{ee'}$ that enters through the resonance $\Omega_1-\Omega_3=\omega_{ee'}-i\gamma_{ee'}$. We indicate the corresponding states rather than transitions, since in the loop all the transitions are calculated using the ground state as reference, thus $e_j\to e_jg$. This follows from the bookkeeping of the wavefunction. We first discuss the left column for which we set the dephasing rate $\gamma_{ee'}=1$ meV. Fig. 3a shows the result for a classical light with narrow intraband dephasing $\gamma_{ee'}=1$ meV.  It gives a diagonal cross peak $e=e'$ and one pair of weak side peaks parallel to the main diagonal at $(e,e')=(e_2,e_3)$. The remaining two pairs of side peaks at $(e,e')= (e_1,e_2)$ and $(e,e')=(e_1,e_3)$ are too weak to be seen. Fig. 3d shows the  signal obtained  using entangled photons  with short entanglement time $T_e=10 $ fs. Panels a and d are very similar. The situation changes as the entanglement time is increased further. For $T_e=100$ fs in Fig. 3g we observe two additional strong side cross peak pairs with $(e,e')=(e_1,e_3)$ and $(e,e')=(e_1,e_2)$. The weak peak at $(e,e')=(e_2,e_3)$ is significantly enhanced as well. Note that the visibility and intensity of the side peaks is enhanced for longer entanglement time. This can be explained as follows: the long entanglement time together with the broad pump bandwidth $\sigma_p$ defines a parameter regime where the entanglement manifests with positive frequency correlation, i.e. the difference between frequencies of entangled photons has a narrow distribution \cite{Shi09}. In this case the narrow resonance occurs for $\Omega_1-\Omega_3$ and the inverse of the entanglement time is an effective bandwidth of the pulse envelope which oscillates as a function of frequency (sinc-function). The oscillating envelope enhances or suppresses certain peaks and the longer entanglement time provides the narrow bandwidth which implies a higher frequency resolution.  The other two columns  in Fig. 3 repeat these calculations for larger dephasing rates $\gamma_{ee'}$. If the intraband dephasing $\gamma_{ee'}$ is broader then the classical result depicted in Fig. 3b shows broadening of the main $e=e'$ peak and  the side peaks are significantly suppressed compared to those shown in Fig. 3a. Further increase of $\gamma_{ee'}$ and use of classical fields leads to further broadening of the main diagonal peak whereas the side peaks completely disappear (see Fig. 3c). The same argument applies to the entangled fields with short dephasing time shown in Fig. 3e-f. Broader dephasing rate covers the side peaks and only the main diagonal peak $e=e'$ remains strong and broad. For long entanglement time, intraband dephasing leads to broadening and enhancement of the side peaks. For instance in Fig. 1g the side peaks at $(e,e')=(e_1,e_3)$ are quite weak. Same peaks are broadened and enhanced in Fig. 3h and even more so in Fig. 3i. Thus, the display $(\Omega_1,\Omega_3)$ in LOP allows for effective determining of the intraband dephasing for distinct pair of  $e$ and $e'$ states even if intraband dephasing is broad. The advantage of having cross peaks compared to diagonal resonances is that they allow to distinguish individual states even if $\omega_{ee'}$ is degenerate for several pairs of states $e$ and $e'$. If interband dephasing $\gamma_{eg}$ which determines the longitudinal dimension of the cross peak is broad, the cross-peaks will remain distinct if properly engineered entangled light is used for probing these states. Note that the above parameter regime is different from the one studied in Ref. \cite{Sch13} where a narrow pump bandwidth and short entanglement time give rise to negative frequency correlations and a narrow sum frequency resonance \cite{Shi09}. That regime will be discussed in Section 6.

 \begin{figure}[t]
\begin{center}
\includegraphics[trim=0cm 0cm 0cm 0cm,angle=0, width=0.9\textwidth]{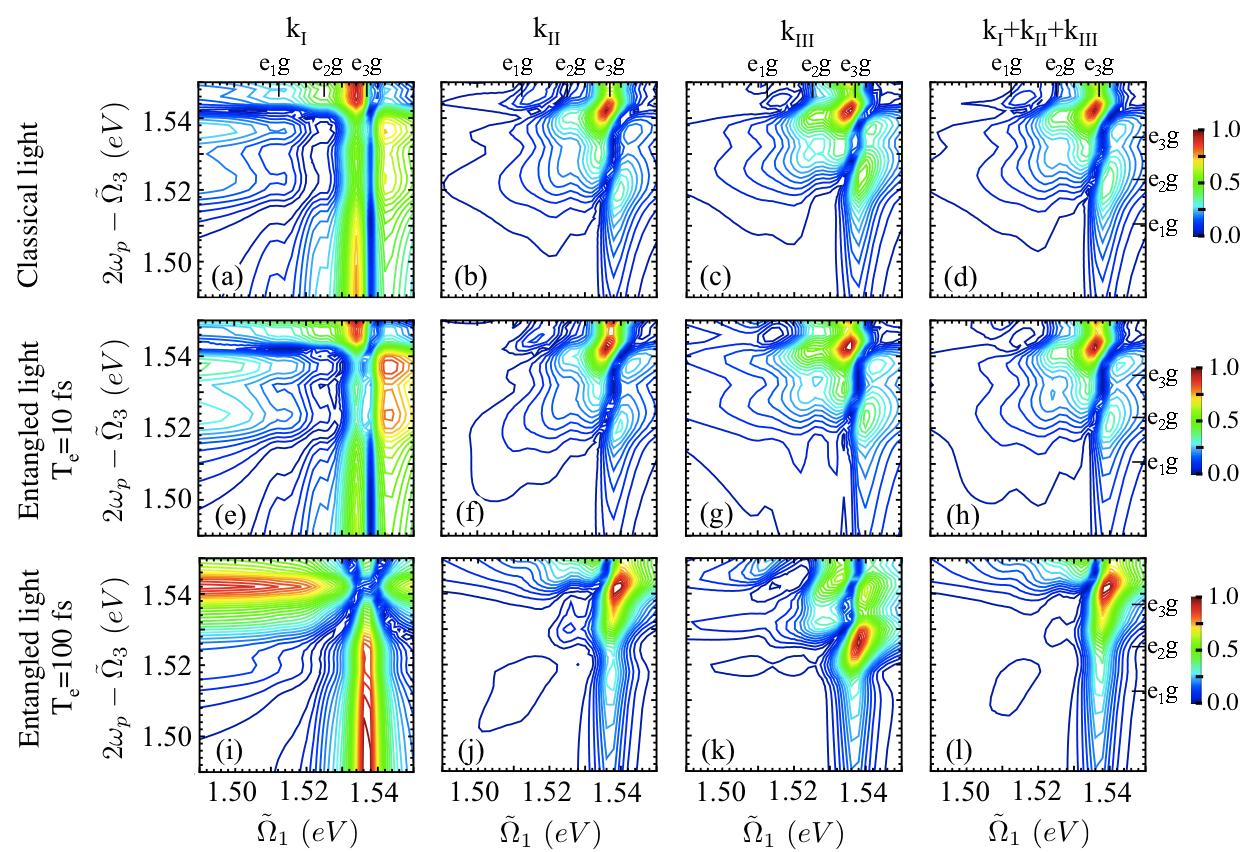}
\end{center}
\caption{(Color online) LAP  $\mathbf{k}_I$, $\mathbf{k}_{II}$, $\mathbf{k}_{III}$ signals Eqs. (\ref{eq:k1}) - (\ref{eq:k3}) $S_{\mathbf{k}_j}(\tilde{\Omega}_1,t_2=0,\tilde{\Omega}_3)$, $j=I,II,III$ for molecular trimer using classical light - top row, entangled light with $T_e=10$ fs - middle row and $T_e=100$ fs - bottom row.  The four columns represent $\mathbf{k}_I$, $\mathbf{k}_{II}$, $\mathbf{k}_{III}$, and $\mathbf{k}_I+\mathbf{k}_{II}+\mathbf{k}_{III}$ as indicated. Intraband dephasing $\gamma_{ee'}=1$ meV. All other parameters are the same as in Fig. 3.}
\label{fig:PP}
\end{figure}

We now turn to interband dephasing. The LOP allows to extract the detailed information about $\gamma_{fe}$. Fig. 4 depicts $S_{LOP}(\Omega_1,\Omega_2,\tau_3=0)$. Fig. 4a shows the signal using classical light at narrow dephasing rate $\gamma_{fe}=1$ meV.  The spectra are dominated by the resonance $\Omega_2-\Omega_1=\omega_{fe}-i\gamma_{fe}$. There are total nine possible transitions between three states $e_j\to f_k$, $j,k=1,2,3$. For the small dephasing rate as in Fig. 4a one can resolve individual cross peaks and extract the information about the interband dephasing. As the dephasing rate is increased, excitation by classical light does not allow to resolve individual transitions but one can rather see only well resolved group of peaks as per Fig. 4b. Further increase the dephasing rate makes the spectra broad and poorly resolved (see Fig. 4c). The short entanglement time used here provides extra selectivity over the distribution of double-excited states via $\Omega_2$ as follows from Fig. 5d. Unlike the classical case where selectivity over $\Omega_2$ and $\Omega_1$ is the same and is determined by the interband dephasing $\gamma_{eg}\sim\gamma_{fg}$, in the entangled case, the time constraint due to $T_e$ provides better selectivity over $\Omega_2$. As the dephasing rate is increased (Fig. 5e) the $\Omega_1$ resolution decreases similarly to the classical case  whereas the selectivity over $\Omega_2$ remains fixed. The same tendency holds if the dephasing is further increased as per Fig. 5f. This allows to resolve individual quantum pathway that contain a single $f$ and single $e$ state and the dephasing $\gamma_{fe}$. Note, that the resolution of $\Omega_2$ is eroded for the longer entanglement time. Therefore the selectivity in both $\Omega_1$ and $\Omega_2$ is eroded quite rapidly with increase of $\gamma_{fe}$ as illustrated in Fig. 5 g-i.

\subsection{LAP signals}

 \begin{figure}[t]
\begin{center}
\includegraphics[trim=0cm 0cm 0cm 0cm,angle=0, width=0.6\textwidth]{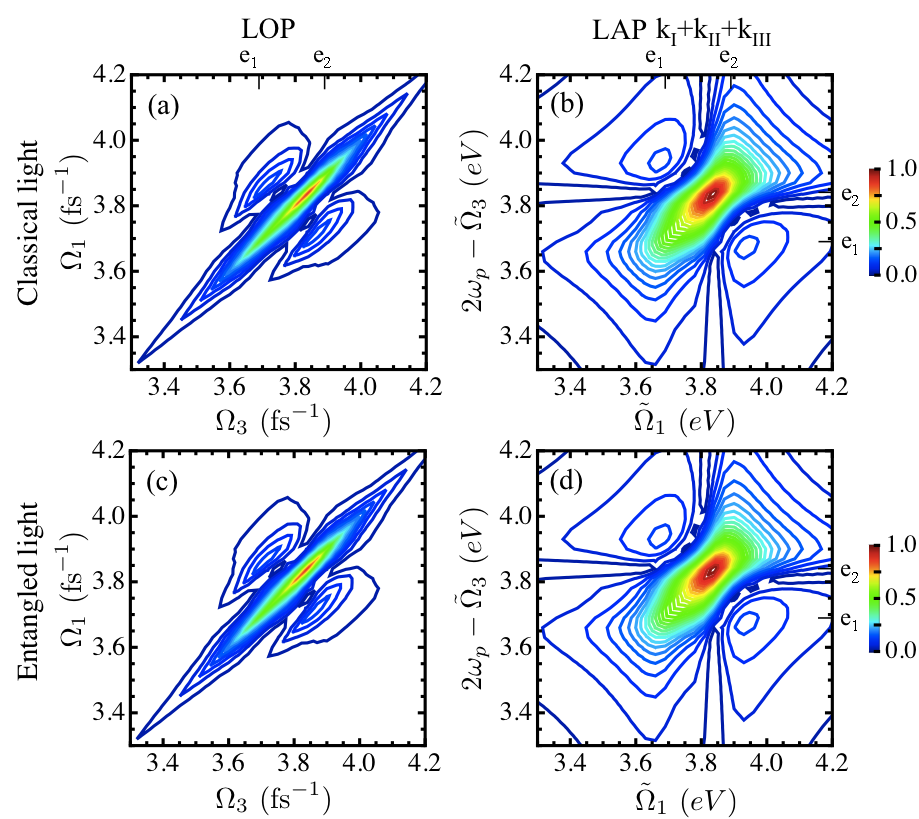}
\end{center}
\caption{(Color online) Left column: $S_{LOP}(\Omega_1,\tau_2=0,\Omega_3)$  Eq. (\ref{eq:S70}) for the molecular dimer model of Ref. \cite{Ray13} calculated using classical light - (a), and entangled light - (c). Right column: same for  $S_{LAP}(\tilde{\Omega}_1,t_2=0,\tilde{\Omega}_3)$ Eq. (\ref{eq:S90}). The dimer has a twist angle 75$^{o}$, coupling strength $+400$ cm$^{-1}$ and monomer transition energy $3.77$ rad fs$^{-1}$. Population relaxation rates  $\gamma_{ee}=\gamma_{e'e'}=0.03$ fs$^{-1}$, dephasing rates $\gamma_{ee'}=0.04$ fs$^{-1}$, $\gamma_{eg}=\gamma_{e'g}=\gamma_{fe}=\gamma_{fe'}=0.08$ fs$^{-1}$, $\gamma_{fg}=0.07$ fs$^{-1}$.}
\label{fig:PP}
\end{figure}

 \begin{figure}[t]
\begin{center}
\includegraphics[trim=0cm 0cm 0cm 0cm,angle=0, width=0.8\textwidth]{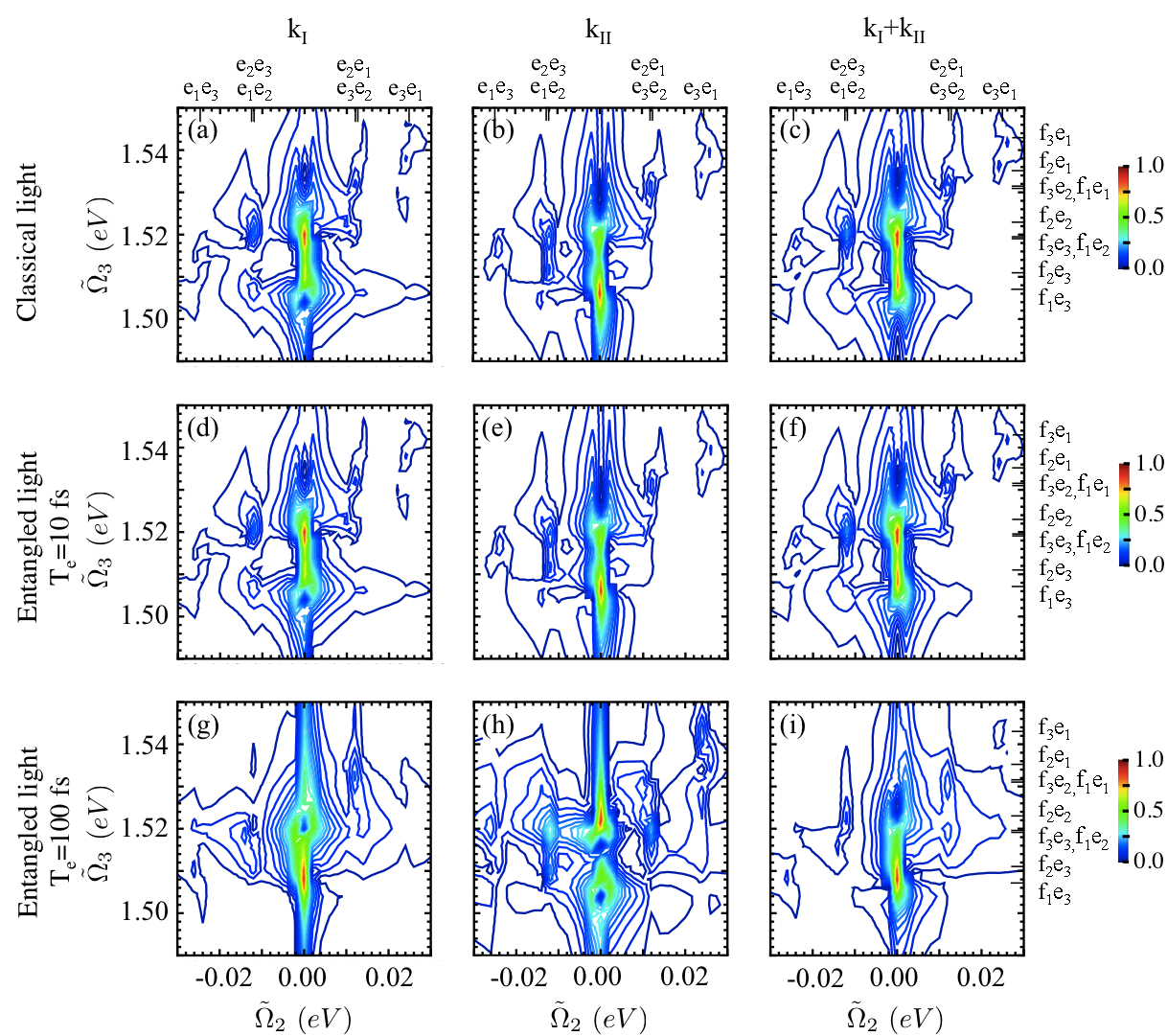}
\end{center}
\caption{(Color online) left column: LAP signal  $S_{\mathbf{k}_I}(t_1=0,\tilde{\Omega}_2,\tilde{\Omega}_3)$ - Eq. (\ref{eq:k1}) for molecular trimer using classical light - top row, entangled light with $T_e=10$ fs - middle row and $T_e=100$ fs - bottom row.  Middle column: same for $\mathbf{k}_{II}$ - Eq. (\ref{eq:k2}), right column: same for the $\mathbf{k}_I+\mathbf{k}_{II}$ signal. The intraband dephasing is $\gamma_{ee'}=1$ meV. All other parameters are the same as in Fig. 3.}
\label{fig:PP}
\end{figure}

As we did for the LOP we show  2D signals obtained by setting one time interval to zero. Fig. 5 depicts the LAP signal $S_{LAP}(\tilde{\Omega}_1,t_2=0,2\omega_p-\tilde{\Omega}_3)$ (\ref{eq:S90})  (we plotted it vs $2\omega_p-\tilde{\Omega}_3$ for a better comparison with Fig. 3). As we did for the LOP we investigate the effect of intraband dephasing $\gamma_{ee'}$. We set $\gamma_{ee'}=1$ meV. Unlike the LOP which contains contributions from all six diagrams in Fig. 2, LAP allows to distinguish between the $\mathbf{k}_I$, $\mathbf{k}_{II}$ and $\mathbf{k}_{III}$ contributions. Fig 5a shows the $\mathbf{k}_I$ signal for the narrow intraband dephasing $\gamma_{ee'}=1$ meV. It is dominated by two resonances for $\tilde{\Omega}_1\simeq\omega_{eg}$, $2\omega_p-\omega_{fe}$, and $\tilde{\Omega}_{3}\simeq \omega_{fe}$, $2\omega_p-\omega_{eg}$. The $\mathbf{k}_{II}$ signal shown in Fig. 5b is dominated by a cross-peak at $\tilde{\Omega}_1+\tilde{\Omega}_3=2\omega_p-\omega_{ee'}-i(\sigma_p+\gamma_{ee'})$. Note, that unlike the LOP, in the case of LAP the width of the $ee'$ resonance is affected by the pump pulse bandwidth $\sigma_p$ and the resonance is broadened as can be seen from Fig. 5b. The same applies to $\mathbf{k}_{III}$. For comparison with the LOP we plot the sum of all three techniques in Fig. 5d. It resembles $\mathbf{k}_{II}$ and $\mathbf{k}_{III}$ and shows that for the same parameters compare to LOP, we get significantly broader resonances and thus, information about intraband dephasing cannot be effectively extracted from this display mode. As shown below it can be done from the $(t_1=0,\tilde{\Omega}_2,\tilde{\Omega}_3)$ display. Unlike the LOP where entanglement at long times $T_e$ plays a crucial role, LAP does not carry extra information about intraband dephasing and essentially gives similar spectra to classical light. Slight changes in peaks intensities can be observed at long entanglement times in $\mathbf{k}_{II}$ and $\mathbf{k}_{III}$ signals (see Fig. 5j,k) compared to short entanglement time in Fig. 5f,g and classical light in Fig. 5 b,c.

It follows from Fig. 3 and Fig. 5 that entanglement is not necessary to reveal the narrow intraband dephasing $\gamma_{ee'}$. The narrow resonances can be observed  with classical light as in Fig. 3d for the right choice of field parameters and if displayed using the LOP. On the other hand the LAP cannot reveal the narrow dephasing neither with nor without entanglement, as shown in Fig. 5. 

Recently 2D spectra of a model dimer with classical and entangled light were calculated in Ref. \cite{Ray13} using a different approach and approximations than used here. Fig. 6 displays  the signals calculated using our approach for the same model dimer parameters of  \cite{Ray13}. The LOP spectra for classical and entangled light are compared in the left column. The corresponding LAP spectra are shown in the right column. We see that entanglement makes no difference in this parameter regime (the two rows are virtually identical). However the scanning protocol does as seen by the two columns. The LOP signals are narrow and clearly resolve the $e_1$ and $e_2$ states whereas the corresponding LAP signals are broad and featureless. 

For more in depth comparison we now describe the signals calculated in  Ref. \cite{Ray13} using our terminology. In that work the entangled LOP spectrum (bottom row of their Fig. 7, our Fig. 6c) was compared with the classical LAP spectrum ( bottom row of their Fig. 6 corresponding to our Fig. 6b). In Ref. \cite{Ray13} the difference was attributed to entanglement effects. Our results show that the difference is solely due to the different scanning protocol (LAP/LOP) and is unrelated to entanglement. Note that the LAP yields three different signals that can be distinguished by the choice of phase, whereas the LOP combines all six contributions into one signal. Furthermore, in order to recover the expressions in Eqs. (21) - (23) and (24) - (29) of Ref. \cite{Ray13} using our model of entangled light we had to take continuous limit for entangled case $\sigma_p\to0$ with $T_{e1}=T_{e2}=0$ and the impulsive limit $\sigma_p\to\infty$ in the case of classical light, which corresponds to two completely different parameter regimes. For a consistent comparison of the classical vs entangled light we used in Fig. 6 the impulsive limit for all four signals. 

The LOP protocol can be generally realized using a pulse shaper as explained earlier. Using our analysis we conclude that the Franson interferometer proposed in Ref. \cite{Ray13} shall provide a convenient method for realizing this protocol experimentally.


 \begin{figure}[t]
\begin{center}
\includegraphics[trim=0cm 0cm 0cm 0cm,angle=0, width=0.8\textwidth]{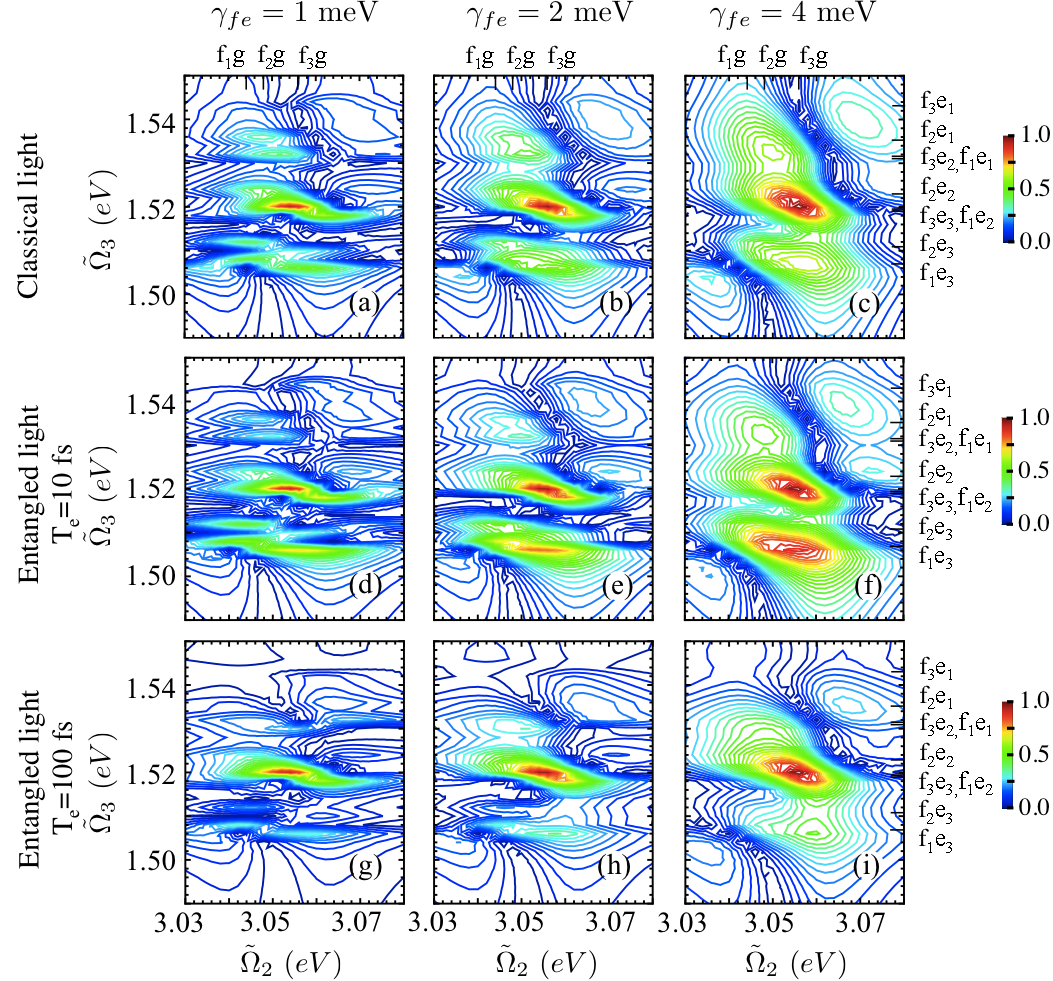}
\end{center}
\caption{(Color online) LAP  $S_{\mathbf{k}_{III}}(t_1=0,\tilde{\Omega}_2,\tilde{\Omega}_3)$ signal Eq. (\ref{eq:k3}) for molecular trimer using classical light - top row, entangled light with $T_e=10$ fs - middle row and $T_e=100$ fs - bottom row. Interband dephasing $\gamma_{fe}=1$ meV - left column, $2$ meV - middle column, and $4$ meV - right column. All other parameters are the same as in Fig. 4.}
\label{fig:PP}
\end{figure}

As demonstrated above, displaying the LAP signal vs $(\tilde{\Omega}_1,t_2=0,\tilde{\Omega}_3)$ does not allow to extract the intraband dephasing $\gamma_{ee'}$ since the spectra are broadened by the pulse bandwidth (see Fig. 5). However we can extract the intraband dephasing by plotting $\mathbf{k}_{I}$ signal - Eq. (\ref{eq:k1}) and $\mathbf{k}_{II}$ - Eq. (\ref{eq:k2}) if displayed vs $(t_1=0,\tilde{\Omega}_2,\tilde{\Omega}_3)$ - see Fig. 7. Note, that here we depicted the ticks along the axes corresponding to the relevant transitions keeping track of the density matrix. The spectra are dominated by $\tilde{\Omega}_2=\omega_{ee'}-i\gamma_{ee'}$ resonance. The $\mathbf{k}_{III}$ signal does not show any features in the vicinity of $\tilde{\Omega}_2=\omega_{ee'}$ so it is not shown. For the narrow dephasing $\gamma_{ee'}$ the spectra of $\mathbf{k}_{I}$ signal  shown in Fig. 7a shows strong diagonal $e=e'$ resonance and weak cross peaks at $(e,e')=(e_1,e_2)$ and $(e,e')=(e_2,e_3)$. The peak at $(e,e')=(e_1,e_3)$ is significantly weaker than the other two. Similar spectra is obtained for $\mathbf{k}_{II}$ signal - Fig. 7b and the total $\mathbf{k}_{I}+\mathbf{k}_{II}$ signal - Fig. 7c. Using entangled light with short entanglement time $T_e=10$ fs, the spectra are virtually identical to the classical light  as shown Fig. 7d-f compared to Fig. 7a-c. The interesting effect occurs for the long entanglement time as in the case of LOP. Fig. 7g shows the side peaks $e\neq e'$ in rephasing signal $\mathbf{k}_{I}$ are suppressed, whereas the nonrephasing contribution $\mathbf{k}_{II}$ in Fig. 7h has enhanced side peaks including $(e,e')=(e_1,e_3)$ resonance that becomes well pronounced. The total $\mathbf{k}_{I}+\mathbf{k}_{II}$ signal depicted in Fig. 7i shows the suppressed side resonances. 

For comparison with the LOP and with  Fig. 4 that  reveals interband dephasing $\gamma_{fe}$ we plot the LAP signal vs $(\tilde{\Omega}_2,\tilde{\Omega}_3)$ in Fig. 8. For a narrow dephasing $\gamma_{fe}=1$ meV the spectra reveals nine $\tilde{\Omega}_3=\omega_{fe}-i\gamma_{fe}$ peaks as shown in Fig. 8a. For broader dephasing $\gamma_{fe}=2$ meV  - Fig. 8b and $\gamma_{fe}=4$ meV - Fig. 8c the spectra are broadened and various peaks overlap. When using entangled light the short entanglement case with $T_e=10$ fs is very similar to the classical light as can be seen by comparing Fig. 8d-f with Fig. 8a-c. Unlike spectra in Figs. 3-7, longer entanglement time does not provide any benefit. Rather it makes various peaks suppressed compared to the classical case as can be seen in Fig. 8g-i. 

\section{Narrowband pulses; Mixed time/frequency-domain scans}

 \begin{figure}[t]
\begin{center}
\includegraphics[trim=0cm 0cm 0cm 0cm,angle=0, width=0.95\textwidth]{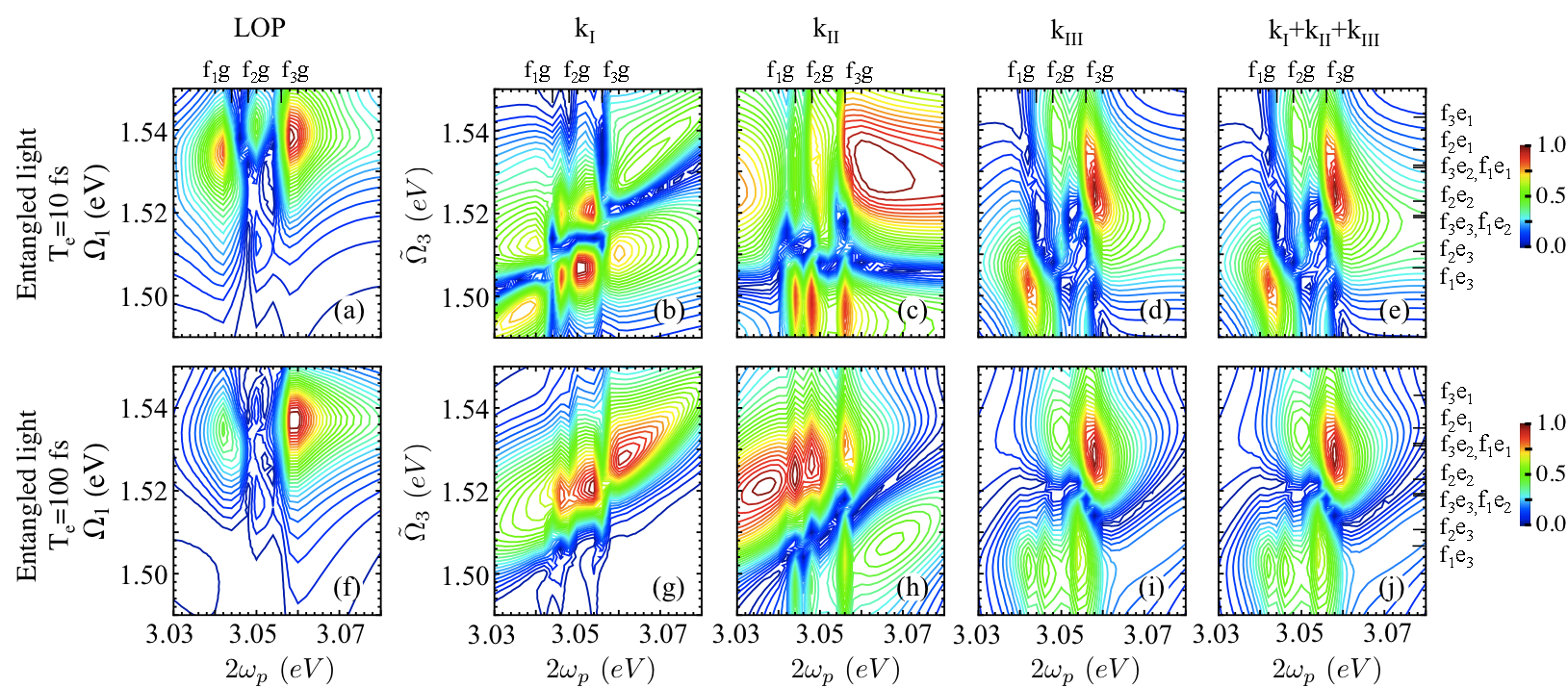}
\end{center}
\caption{(Color online) Top row: $S_{LOP}(\Omega_1,\Omega_2=3.11 eV, \tau_3=0;\omega_p)$ Eq. (\ref{eq:S70}) for entangled light with $T_e=10$ fs - (a), LAP $S_{\mathbf{k}_I}(t_1=0,\tilde{\Omega}_2=3.11 eV,\tilde{\Omega}_3;\omega_p)$ Eq.  (\ref{eq:k1}) - (b), same for $\mathbf{k}_{II}$ signal Eq.  (\ref{eq:k2}) - (c), same for $\mathbf{k}_{III}$ signal Eq.  (\ref{eq:k3}) - (d), same for the $\mathbf{k}_{I}+\mathbf{k}_{II}+\mathbf{k}_{III}$ signal - (e). Bottom row: same as the top row but for $T_e=100$ fs. The pump bandwidth is $\sigma_p=0.8$ meV, interband dephasing $\gamma_{fg}=2$ meV. All other parameters  are the same as in Fig. 4.}
\label{fig:PP}
\end{figure}

So far we investigated multidimensional signals obtained by scanning various time delays between pulses. This time-domain protocol makes sense if the pulses that interact with the system are relatively short. For entangled light this implies that frequencies of the modes corresponding to the twin photons are positively correlated \cite{Bre13}. We demonstrated that this is crucial especially for the long entanglement time where the narrow difference-frequency-resonances can be observed in the spectra of $S_{LOP}(\Omega_1,\tau_2=0,\Omega_3)$, $S_{LAP}(\tilde{\Omega}_1,t_2=0,\tilde{\Omega}_3)$, $S_{LOP}(\Omega_1,\Omega_2,\tau_3=0)$, and $S_{LAP}(t_1=0,\tilde{\Omega}_2,\tilde{\Omega}_3)$ signals. 

In our recent work \cite{Sch13} we have investigated the effects of entanglement on the control of the transport properties in molecular aggregates. Narrow $fg$ resonances were observed when the entangled pair has been generated by narrowband pump and entanglement time is short. In this case narrow pump along with short entanglement time implies negative frequency correlation (the sum of two frequencies is narrowly distributed). This is a different parameter regime than used in Section 5. In the following we consider narrowband pump pulse and $fg$ resonances with entangled photons. In this case we can adopt mixed time-and-frequency domain scanning, where we scan one time delay between pulses and the pump frequency $\omega_p$. Again we compare the LOP and LAP protocols. 

Fig. 9 depicts the corresponding time-and-frequency domain signal. Fig. 9a shows the signal $S_{LOP}(\Omega_1,\Omega_2=3.11 eV, \tau_3=0,\omega_p)$ LOP signal which contains three distinct peaks corresponding to $2\omega_p=\omega_{fg}-i\gamma_{fg}$ resonances for short entanglement time $T_e=10$ fs. As the entanglement time is increased, the peaks become weaker as shown in Fig. 9f. For comparison we depict the corresponding series of LAP signals $S_{\mathbf{k}_j}(t_1=0,\tilde{\Omega}_2=3.11 eV, \tilde{\Omega}_3;\omega_p)$ where $j=I, II, III$. For short entanglement time the $\mathbf{k}_{II}$ signal contains three well pronounced narrow peaks similar to LOP as seen from Fig. 9c, whereas resonances in $\mathbf{k}_I$ and $\mathbf{k}_{III}$ signals are not as clearly seen. For longer entanglement time, all sharp features of LAP spectra become fuzzy (see FIg. 9 g-j) and even in the case of $\mathbf{k}_{II}$, $\omega_{fg}$ resonances become suppressed. This is consistent with earlier results for narrowband pump pulse \cite{Sch13}. 


\section{Conclusions}

Multidimensional optical signals are obtained by subjecting the system to sequences of  short pulses  and generating and analyzing correlation plots between different resonances generated during controlled delay periods. These allow to visualize such an event as a e.g. cross-peak in the space of two frequency variables that are related to Fourier transform of two different delay intervals. Most commonly, the delays are between consecutive chronologically-ordered pulses that can differ by their frequencies, polarizations and wavevectors. Such signals can be naturally described by the density matrix and represented by ladder diagrams. We had  presented a new protocol based on the wavefunction description  that involves  both forward and backward time evolution. This protocol uses different types of delays  represented by loop diagrams and can be realized experimentally by phase cycling.  This  new type of bookkeeping of field-matter interactions that is not based on chronologically time ordered pulses  suggests  a new way of monitoring and displaying various resonances . We demonstrated it for two photon absorption experiments with incoherent fluorescence detection in a molecular aggregate with classical and entangled light.

Broadband entangled light with long entanglement time allows to selectively reduce the background and reveal certain resonances because of intrinsic frequency correlations due to entanglement. The resonances remain well resolved even for the short dephasing which typically is a source of strong background for the signals measured with classical fields. In particular, entangled light and the loop-based protocol can reveal intra and interband dephasing in the single and double exciton manifold  not possible by classical light. We demonstrated better-resolved signals  compared to those obtained with  standard ladder scanning protocol. Entangled light causes correlations of the various time delay variables  thus providing new  spectroscopic windows  and physical picture of the  system dynamics. The current formalism can be readily applied  for an arbitrary state of light including stochastic, squeezed or  other quantum and classical states. The signals are  given by sums of products of  four-point correlation functions of the electric field  and matter which can be calculated for  arbitrary pulse shapes and bandwidths including temporally overlapping pulses. The necessary Liouville space Green's functions can be evaluated by taking bath effects into account, e.g. pure dephasing, inhomogeneous broadening, transport and other dynamical bath effects.

\appendix

\section{~~~~~~~~~~~~~Time-domain signals using LOP with entangled photons}
Here we evaluate the frequency integrals in Eq. (\ref{eq:S60}). The time-domain LOP signal then reads
\begin{align}
S_{LOP}^{(1)}&(\tau_1,\tau_2,\tau_3)=\mathcal{I}\frac{1}{2\sigma_p\hbar^4}\sum_{e,e'}\mu_{ge'}^a\mu_{e'f}^b\mu_{fe}^{c*}\mu_{eg}^{d*}\theta(\tau_2)\mathcal{G}_{ef}(\tau_1)\mathcal{G}_{ge'}(\tau_3)\notag\\
\times&\left[\theta(\tau_2-\tau_1)e^{[2i\omega_p-\sigma_p](\tau_2-\tau_1)}\mathcal{G}_{gf}(-2\omega_p-i\sigma_p)\mathcal{F}(2\omega_p-\omega_{fe}-i(\gamma_{fe}-\sigma_p),\omega_{fe}+i\gamma_{fe},\omega_{e'g}+i\gamma_{e'g})\right.\notag\\
&\left.+\theta(\tau_1-\tau_2)e^{-[2i\omega_p+\sigma_p](\tau_1-\tau_2)}\mathcal{G}_{gf}(-2\omega_p+i\sigma_p)\mathcal{F}(2\omega_p-\omega_{fe}-i(\gamma_{fe}+\sigma_p),\omega_{fe}+i\gamma_{fe},\omega_{e'g}+i\gamma_{e'g})\right.\notag\\
&\left.+2\sigma_p\mathcal{G}_{gf}(\tau_2-\tau_1)\mathcal{G}_{gf}(-2\omega_p-i\sigma_p)\mathcal{G}_{gf}(-2\omega_p+i\sigma_p)\mathcal{F}(\omega_{eg}-i\gamma_{eg},\omega_{fe}+i\gamma_{fe},\omega_{e'g}+i\gamma_{e'g})\right],
\end{align}
\begin{align}
S_{LOP}^{(2)}&(\tau_1,\tau_2,\tau_3)=-\mathcal{I}\frac{1}{2\sigma_p\hbar^4}\sum_{e,e'}\mu_{ge'}^a\mu_{e'f}^b\mu_{fe}^{c*}\mu_{eg}^{d*}\theta(\tau_1)\theta(\tau_2)\mathcal{G}_{ee'}(\tau_3)\notag\\
\times&\left[\theta(\tau_2-\tau_1+\tau_3)e^{[2i\omega_p-\sigma_p](\tau_2-\tau_1+\tau_3)}\mathcal{G}_{ef}(\tau_1-\tau_3)\mathcal{G}_{gf}(-2\omega_p-i\sigma_p)\right.\notag\\
&\left.\times\mathcal{F}(2\omega_p-\omega_{fe}-i(\gamma_{fe}-\sigma_p),\omega_{fe}+i\gamma_{fe},2\omega_p-\omega_{fe'}-i(\gamma_{fg}-\gamma_{e'g}-\sigma_p))\right.\notag\\
&\left.+\theta(\tau_1-\tau_3-\tau_2)e^{-[2i\omega_p+\sigma_p](\tau_1-\tau_3+\tau_2)}\mathcal{G}_{ef}(\tau_1-\tau_3)\mathcal{G}_{gf}(-2\omega_p+i\sigma_p)\right.\notag\\
&\left.\times\mathcal{F}(2\omega_p-\omega_{fe}-i(\gamma_{fe}+\sigma_p),\omega_{fe}+i\gamma_{fe},2\omega_p-\omega_{fe'}-i(\gamma_{fg}-\gamma_{e'g}+\sigma_p))\right.\notag\\
&\left.-\theta(\tau_2)e^{[2i\omega_p-\sigma_p]\tau_2}\mathcal{G}_{eg}(\tau_1-\tau_3)\mathcal{G}_{gf}(-2\omega_p-i\sigma_p)\right.\notag\\
&\left.\times\mathcal{F}(\omega_{eg}-i\gamma_{eg},2\omega_p-\omega_{eg}+i(\gamma_{eg}+\sigma_p),\omega_{e'g}+i\gamma_{e'g})\right.\notag\\
&\left.+2\sigma_p[\mathcal{G}_{ef}(\tau_1-\tau_3)\mathcal{G}_{gf}(\tau_2-\tau_1+\tau_3)-\mathcal{G}_{eg}(\tau_1-\tau_3)\mathcal{G}_{gf}(\tau_2)]\right.\notag\\
&\left.\times\mathcal{G}_{gf}(-2\omega_p-i\sigma_p)\mathcal{G}_{gf}(-2\omega_p+i\sigma_p)\mathcal{F}(\omega_{eg}-i\gamma_{eg},\omega_{fe}+i\gamma_{fe},\omega_{e'g}+i\gamma_{e'g})\right],
\end{align}
\begin{align}
S_{LOP}^{(3)}&(\tau_1,\tau_2,\tau_3)=\mathcal{I}\frac{1}{2\sigma_p\hbar^4}\sum_{e,e'}\mu_{ge'}^a\mu_{e'f}^b\mu_{fe}^{c*}\mu_{eg}^{d*}\theta(\tau_1)\theta(\tau_2)\theta(\tau_3)\notag\\
\times&\left[-\theta(\tau_1-\tau_2)e^{-[2i\omega_p+\sigma_p](\tau_1-\tau_2)}\mathcal{G}_{ef}(\tau_1)\mathcal{G}_{ge'}(\tau_3)\mathcal{G}_{gf}(-2\omega_p+i\sigma_p)\right.\notag\\
&\left.\times\mathcal{F}(2\omega_p-\omega_{fe}-i(\gamma_{fe}+\sigma_p),\omega_{fe}+i\gamma_{fe},\omega_{e'g}+i\gamma_{e'g})\right.\notag\\
&\left.-\theta(\tau_2-\tau_1)e^{[2i\omega_p-\sigma_p](\tau_2-\tau_1)}\mathcal{G}_{ef}(\tau_1)\mathcal{G}_{ge'}(\tau_3)\mathcal{G}_{gf}(-2\omega_p-i\sigma_p)\right.\notag\\
&\left.\times\mathcal{F}(2\omega_p-\omega_{fe}-i(\gamma_{fe}-\sigma_p),\omega_{fe}+i\gamma_{fe},\omega_{e'g}+i\gamma_{e'g})\right.\notag\\
&\left.-2\sigma_p\mathcal{G}_{gf}(\tau_2-\tau_1)\mathcal{G}_{ef}(\tau_1)\mathcal{G}_{ge'}(\tau_3)\mathcal{G}_{gf}(-2\omega_p-i\sigma_p)\mathcal{G}_{gf}(-2\omega_p+i\sigma_p)\right.\notag\\
&\left.\times\mathcal{F}(\omega_{eg}-i\gamma_{eg},\omega_{fe}+i\gamma_{fe},\omega_{e'g}+i\gamma_{e'g})\right.\notag\\
&\left.+\theta(\tau_2)e^{[2i\omega_p-\sigma_p]\tau_2}\mathcal{G}_{eg}(\tau_1)\mathcal{G}_{ge'}(\tau_3)\mathcal{G}_{gf}(-2\omega_p-i\sigma_p)\right.\notag\\
&\left.\times\mathcal{F}(\omega_{eg}-i\gamma_{eg},2\omega_p-\omega_{eg}+i(\sigma_p+\gamma_{eg}),\omega_{e'g}+i\gamma_{e'g})\right.\notag\\
&\left.+2\sigma_p\mathcal{G}_{gf}(\tau_2)\mathcal{G}_{eg}(\tau_1)\mathcal{G}_{ge'}(\tau_3)\mathcal{G}_{gf}(-2\omega_p-i\sigma_p)\mathcal{G}_{gf}(-2\omega_p+i\sigma_p)\right.\notag\\
&\left.\times\mathcal{F}(\omega_{eg}-i\gamma_{eg},\omega_{fe}+i\gamma_{fe},\omega_{e'g}+i\gamma_{e'g})\right.\notag\\
&\left.+\theta(\tau_1-\tau_3-\tau_2)e^{-[2i\omega_p+\sigma_p](\tau_1-\tau_3-\tau_2)}\mathcal{G}_{ef}(\tau_1-\tau_3)\mathcal{G}_{ee'}(\tau_3)\mathcal{G}_{gf}(-2\omega_p+i\sigma_p)\right.\notag\\
&\left.\times\mathcal{F}(2\omega_p-\omega_{fe}-i(\gamma_{fe}+\sigma_p),\omega_{fe}+i\gamma_{fe},2\omega_p-\omega_{fe'}-i(\gamma_{fg}-\gamma_{e'g}+\sigma_p))\right.\notag\\
&\left.+\theta(\tau_2+\tau_3-\tau_1)e^{[2i\omega_p-\sigma_p](\tau_2+\tau_3-\tau_1)}\mathcal{G}_{ef}(\tau_1-\tau_3)\mathcal{G}_{ee'}(\tau_3)\mathcal{G}_{gf}(-2\omega_p-i\sigma_p)\right.\notag\\
&\left.\times\mathcal{F}(2\omega_p-\omega_{fe}-i(\gamma_{fe}-\sigma_p),\omega_{fe}+i\gamma_{fe},2\omega_p-\omega_{fe'}-i(\gamma_{fg}-\gamma_{e'g}-\sigma_p))\right.\notag\\
&\left.+2\sigma_p\mathcal{G}_{gf}(\tau_2+\tau_3-\tau_1)\mathcal{G}_{ef}(\tau_1-\tau_3)\mathcal{G}_{ee'}(\tau_3)\mathcal{G}_{gf}(-2\omega_p-i\sigma_p)\mathcal{G}_{gf}(-2\omega_p+i\sigma_p)\right.\notag\\
&\left.\times\mathcal{F}(\omega_{eg}-i\gamma_{eg},\omega_{fe}+i\gamma_{fe},\omega_{e'g}+i\gamma_{e'g})\right.\notag\\
&\left.-\theta(\tau_2)e^{[2i\omega_p-\sigma_p]\tau_2}\mathcal{G}_{eg}(\tau_1-\tau_3)\mathcal{G}_{ee'}(\tau_3)\mathcal{G}_{gf}(-2\omega_p-i\sigma_p)\right.\notag\\
&\left.\times\mathcal{F}(\omega_{eg}-i\gamma_{eg},2\omega_p-\omega_{eg}+i(\gamma_{eg}+\sigma_p),\omega_{e'g}+i\gamma_{e'g})\right.\notag\\
&\left.-2\sigma_p\mathcal{G}_{gf}(\tau_2)\mathcal{G}_{eg}(\tau_1-\tau_3)\mathcal{G}_{ee'}(\tau_3)\mathcal{G}_{gf}(-2\omega_p-i\sigma_p)\mathcal{G}_{gf}(-2\omega_p+i\sigma_p)\right.\notag\\
&\left.\times\mathcal{F}(\omega_{eg}-i\gamma_{eg},\omega_{fe}+i\gamma_{fe},\omega_{e'g}+i\gamma_{e'g})\right],
\end{align}
\begin{align}
S_{LOP}^{(4)}(\tau_1,\tau_2,\tau_3)=S^{(1)*}(\tau_3,\tau_2,\tau_1),
\end{align}
\begin{align}
S_{LOP}^{(5)}(\tau_1,\tau_2,\tau_3)=S^{(2)*}(\tau_3,\tau_2,\tau_1),
\end{align}
\begin{align}
S_{LOP}^{(6)}(\tau_1,\tau_2,\tau_3)=S^{(3)*}(\tau_3,\tau_2,\tau_1),
\end{align}
where
\begin{align}
\mathcal{F}(\omega_a,\omega_b,\omega_d)=\tilde{\Phi}(\omega_a,\omega_b)\tilde{\Phi}^{*}(\omega_a+\omega_b-\omega_d,\omega_d).
\end{align}
and
\begin{align}
\Phi(\omega_a,\omega_b)=A_p(\omega_a+\omega_b)\tilde{\Phi}(\omega_a,\omega_b).
\end{align}

\section{~~~~~~~~~~~~~Time-domain signals using LAP with entangled photons}

Evaluating the frequency integrals in (\ref{eq:S80}) we obtain for the time-domain LAP signal
\begin{align}
S_{LAP}^{(1)}&(t_1,t_2,t_3)=\mathcal{I}\frac{1}{2\sigma_p\hbar^4}\sum_{e,e'}\mu_{ge'}^a\mu_{e'f}^b\mu_{fe}^{c*}\mu_{eg}^{d*}\mathcal{G}_{ef}(t_3)\mathcal{G}_{ge'}(t_1)\notag\\
\times&\left[\theta(t_2)e^{[2i\omega_p-\sigma_p]t_2}\mathcal{G}_{gf}(-2\omega_p-i\sigma_p)\mathcal{F}(2\omega_p-\omega_{fe}-i(\gamma_{fe}-\sigma_p),\omega_{fe}+i\gamma_{fe},\omega_{e'g}+i\gamma_{e'g})\right.\notag\\
&\left.+2\sigma_p\mathcal{G}_{gf}(t_2)\mathcal{G}_{gf}(-2\omega_p-i\sigma_p)\mathcal{G}_{gf}(-2\omega_p+i\sigma_p)\mathcal{F}(\omega_{eg}-i\gamma_{eg},\omega_{fe}+i\gamma_{fe},\omega_{e'g}+i\gamma_{e'g})\right],
\end{align}
\begin{align}
S_{LAP}^{(2)}&(t_1,t_2,t_3)=-\mathcal{R}\frac{1}{2\sigma_p\hbar^4}\sum_{e,e'}\mu_{ge'}^a\mu_{e'f}^b\mu_{fe}^{c*}\mu_{eg}^{d*}\mathcal{G}_{ee'}(t_2)\notag\\
\times&\left[e^{[2i\omega_p-\sigma_p]t_3}\mathcal{G}_{eg}(t_3)\mathcal{G}_{eg}(t_1)\mathcal{G}_{gf}(-2\omega_p-i\sigma_p)\right.\notag\\
&\left.\times\mathcal{F}(\omega_{eg}-i\gamma_{eg},2\omega_p-\omega_{eg}+i(\sigma_p+\gamma_{eg}),\omega_{e'g}+i\gamma_{e'g})\right.\notag\\
&\left.+e^{-[2i\omega_p+\sigma_p]t_1}\mathcal{G}_{ef}(t_3)\mathcal{G}_{ef}(t_1)\mathcal{G}_{gf}(-2\omega_p+i\sigma_p)\right.\notag\\
&\left.\times\mathcal{F}(2\omega_p-\omega_{fe}-i(\gamma_{fe}+\sigma_p),\omega_{fe}+i\gamma_{fe},2\omega_p-\omega_{fe'}-i(\gamma_{fg}-\gamma_{e'g}+\sigma_p))\right.\notag\\
&\left.+2i\sigma_p\mathcal{G}_{ef}(t_3)\mathcal{G}_{eg}(t_1)\mathcal{G}_{gf}(-2\omega_p-i\sigma_p)\mathcal{G}_{gf}(-2\omega_p+i\sigma_p)\right.\notag\\
&\left.\times\mathcal{F}(\omega_{eg}-i\gamma_{eg},\omega_{fe}+i\gamma_{fe},\omega_{e'g}+i\gamma_{e'g})\right],
\end{align}
\begin{align}
S_{LAP}^{(3)}&(t_1,t_2,t_3)=\mathcal{R}\frac{1}{2\sigma_p\hbar^4}\sum_{e,e'}\mu_{ge'}^a\mu_{e'f}^b\mu_{fe}^{c*}\mu_{eg}^{d*}\theta(t_1)\theta(t_2)\theta(t_3)\notag\\
\times&\left[-e^{[2i\omega_p-\sigma_p]t_3}\mathcal{G}_{eg}(t_3)\mathcal{G}_{eg}(t_2)\mathcal{G}_{ge'}(t_2+t_1)\mathcal{G}_{gf}(-2\omega_p-i\sigma_p)\right.\notag\\
&\left.\times\mathcal{F}(\omega_{eg}-i\gamma_{eg},2\omega_p-\omega_{eg}+i(\sigma_p+\gamma_{eg}),\omega_{e'g}+i\gamma_{e'g})\right.\notag\\
&\left.-e^{-[2i\omega_p+\Gamma]t_2}\mathcal{G}_{ef}(t_3)\mathcal{G}_{ef}(t_2)\mathcal{G}_{ge'}(t_2+t_1))\mathcal{G}_{gf}(-2\omega_p+i\sigma_p)\right.\notag\\
&\left.\times\mathcal{F}(2\omega_p-\omega_{fe}-i(\gamma_{fe}+\sigma_p),\omega_{fe}+i\gamma_{fe},\omega_{e'g}+i\gamma_{e'g})\right.\notag\\
&\left.-2i\sigma_p\mathcal{G}_{ef}(t_3)\mathcal{G}_{eg}(t_2)\mathcal{G}_{ge'}(t_2+t_1)\mathcal{G}_{gf}(-2\omega_p-i\sigma_p)\mathcal{G}_{gf}(-2\omega_p+i\sigma_p)\right.\notag\\
&\left.\times \mathcal{F}(\omega_{eg}-i\gamma_{eg},\omega_{fe}+i\gamma_{fe},\omega_{e'g}+i\gamma_{e'g})\right.\notag\\
&\left.+ie^{(2i\omega_p-\sigma_p]t_3}\mathcal{G}_{eg}(t_3-t_1)\mathcal{G}_{ee'}(t_2+t_1)\mathcal{G}_{gf}(-2\omega_p-i\sigma_p)\right.\notag\\
&\left.\times\mathcal{F}(\omega_{eg}-i\gamma_{eg},2\omega_p-\omega_{eg}+i(\sigma_p+\gamma_{eg}),\omega_{e'g}+i\gamma_{e'g})\right.\notag\\
&\left. +\theta(t_1-t_3)e^{[2i\omega_p-\sigma_p]t_1-[i\omega_{fe}-\gamma_{fe}](t_1-t_3)}\mathcal{G}_{ee'}(t_2+t_1)\mathcal{G}_{gf}(-2\omega_p-i\sigma_p)\right.\notag\\
&\left.\times\mathcal{F}(2\omega_p-\omega_{fe}-i(\gamma_{fe}-\sigma_p),\omega_{fe}+i\gamma_{fe},2\omega_p-\omega_{fe'}-i(\gamma_{fg}-\gamma_{e'g}-\sigma_p)\right.\notag\\
&\left.-ie^{[2i\omega_p-\sigma_p]t_1-[i\omega_{fe}-\gamma_{fe}]t_1}\mathcal{G}_{ef}(t_3)\mathcal{G}_{ee'}(t_2+t_1)\mathcal{G}_{gf}(-2\omega_p+i\sigma_p)\right.\notag\\
&\left.\times\mathcal{F}(2\omega_p-\omega_{fe}-i(\gamma_{fe}-\sigma_p),\omega_{fe}+i\gamma_{fe},2\omega_p-\omega_{fe'}-i(\gamma_{fg}-\gamma_{e'g}-\sigma_p)\right],
\end{align}
\begin{align}
S_{LAP}^{(4)}(t_1,t_2,t_3)=S^{(1)*}(t_1,t_2,t_3),
\end{align}
\begin{align}
S_{LAP}^{(5)}(t_1,t_2,t_3)=S^{(2)*}(t_1,t_2,t_3),
\end{align}
\begin{align}
S_{LAP}^{(6)}(t_1,t_2,t_3)=S^{(3)*}(t_1,t_2,t_3).
\end{align}

\ack
We wish to thank Frank Schlawin for stimulating discussions and gratefully acknowledge the support of the National Science Foundation through Grant No. CHE-1058791 and computations are supported by CHE-0840513,  the Chemical Sciences, Geosciences and Biosciences Division, Office of Basic Energy Sciences, Office of Science, US Department of Energy and the National Institute of Health Grant No. GM-59230.

\section*{References}



\end{document}